%% file: 0_lpac-paper_arXiv.tex
\pdfoutput=1

\documentclass[a4paper,9pt,twocolumn]{article}

\usepackage{geometry}                
\usepackage{graphicx}
\usepackage{dcolumn}    
\usepackage{bm}			

\usepackage[utf8]{inputenc}
\usepackage[T1]{fontenc}
\usepackage{mathptmx}	
\usepackage{etoolbox}

\usepackage[table]{xcolor}	
\usepackage{amsmath}	
\usepackage{amssymb} 	
\usepackage{amsthm}		
\usepackage{mathtools}
\usepackage{microtype}  
\usepackage{ifthen}		
\usepackage{bbm}		
\usepackage{hyperref}	
\hypersetup{pdfauthor={Bianucci, T. and Zechner, C.},pdftitle={A local polynomial moment approximation for compartmentalised biochemical systems}}
\usepackage{cleveref}	
\usepackage[font=footnotesize,labelfont=bf]{caption}	
\usepackage{xfrac}		

\usepackage{tabularx}
\newcolumntype{Y}{>{\centering\arraybackslash}X}
\usepackage{multirow}
\usepackage{cellspace} 	

\usepackage{todonotes} 	
\usepackage{soul} 		

\usepackage[affil-it]{authblk}	

\makeatletter
\renewenvironment{abstract}{%
	\small\bfseries
	\paragraph{\normalsize\itshape\abstractname:}
	}%
	{\if@twocolumn\else\par\bigskip\fi}
\makeatother

\usepackage{sectsty}
\sectionfont{\fontsize{11}{10}\selectfont}
\subsectionfont{\fontsize{10}{10}\selectfont}
\paragraphfont{\fontsize{9}{10}\selectfont}

\input{1_definitions}		

\title{A local polynomial moment approximation for compartmentalised biochemical systems}
\author[1,2]{Tommaso Bianucci}
\author[1,2,3,*]{Christoph Zechner}

\affil[1]{Max Planck Institute of Molecular Cell Biology and Genetics, Pfotenhauerstraße 108, 01307 Dresden, Germany}
\affil[2]{Center for Systems Biology Dresden, Pfotenhauerstraße 108, 01307 Dresden, Germany}
\affil[2]{Cluster of Excellence Physics of Life, TU Dresden, Arnoldstraße 18, 01307 Dresden, Germany}
\affil[*]{Correspondence to: zechner@mpi-cbg.de}
\date{June 15, 2023}

\begin{document}
\maketitle

\begin{abstract}%
Compartmentalised biochemical reactions are a ubiquitous building block of biological systems. The interplay between chemical and compartmental dynamics can drive rich and complex dynamical behaviors that are difficult to analyse mathematically -- especially in the presence of stochasticity. We have recently proposed an effective moment equation approach to study the statistical properties of compartmentalised biochemical systems. So far, however, this approach is limited to polynomial rate laws and moreover, it relies on suitable moment closure approximations, which can be difficult to find in practice. In this work we propose a systematic method to derive closed moment dynamics for compartmentalised biochemical systems. We show that for the considered class of systems, the moment equations involve expectations over functions that factorize into two parts, one depending on the molecular content of the compartments and one depending on the compartment number distribution. Our method exploits this structure and approximates each function with suitable polynomial expansions, leading to a closed system of moment equations. We demonstrate the method using three systems inspired by cell populations and organelle networks and study its accuracy across different dynamical regimes.%
\end{abstract}

\section{Introduction}
Compartmentalisation of biochemical reactions is one of the most characteristic and fundamental
properties of living systems. It occurs hierarchically across length scales,
ranging from organs and tissues, down to cells and
subcellular structures.
Compartmentalisation provides 
functional segregation in space and time and allows for modularity in the assembly of larger scale structure and function.

Compartmentalised reaction systems
can be described as populations of 
enclosed volumes (e.g., organelles or cells)
that evolve with time. 
Individual compartments contain molecules that can undergo chemical reactions, 
while the compartments themselves can exhibit temporal dynamics as well. 
As an example, two compartments may fuse or divide, or new compartments may be created 
or exit the system as observed in organelle networks for instance~\cite{Alberts2008MolecularBO}. 
The combination of chemical reactions and compartment dynamics results in challenging 
multiscale problems, which attracted substantial interest in the past. 

Population balance equations (PBEs) provide a general mathematical framework to 
describe a broad class of compartmentalized reaction systems~\cite{ramkrishna2014population,ramkrishna2000population}. 
In the context of biology, they have been applied to 
endosomal networks~\cite{foret2012general}, 
heterogeneous cell populations~\cite{henson2003dynamic,waldherr2018estimation} 
and clone-size dynamics~\cite{rulands2018universality} 
to name a few. 
PBEs are typically formulated in the thermodynamic limit and as such apply to 
systems where fluctuations at both the chemical and compartmental level are 
negligible. 
While stochastic systems have also been considered within the population balance 
framework~\cite{ramkrishna1973puristic,ramkrishna2000population}, they are typically accesible only through computationally expensive
 Monte Carlo simulation.

In recent years, there has been increased interest in 
the theoretical and computational analysis 
of compartmentalized reaction systems in the presence of stochasticity
~\cite{thomas2019intrinsic,Duso2020,lunz2022revisiting,duso2021shared,bansaye2015some,anderson2023stochastic}. 
In contrast to bulk chemical systems, stochasticity arises at two levels in these systems. 
First, chemical reactions inside individual compartments take place at random times, 
leading to random variations in each compartment's state~\cite{Gillespie2007}. 
Second, compartment events such as intake, division or turnover exhibit 
stochasticity such that the population size fluctuates with time. 
Most existing computational approaches to analyze compartmentalized reaction systems 
focus on models of cell populations, where intracellular processes are combined 
with population-level dynamics such as cell growth, division or phenotypic 
selection~\cite{jia2021cell,thomas2019intrinsic,thomas2021coordination,lunz2022revisiting,aditya2022using}. 
Since the number of cells is typically large in these systems, one can consider 
the limit of infinitely large populations, where fluctuations in the 
population-level dynamics become negligible. 
An important consequence of this limit is that the entire population can be 
described by a single probability density function $p(x, t)$, which captures how 
the compartment state is distributed across the population at time $t$. 
The dynamics of $p(x, t)$ can then be derived and solved using direct numerical 
integration~\cite{lunz2021beyond,munsky2006finite} 
or moment approximation techniques~\cite{lunz2022revisiting}.

The situation becomes more complex when the number of 
compartments is small, such that the limit of large populations is no longer applicable. 
This can be the case for organelle networks inside cells, or small cell communities 
at early developmental stages. 
When the population size is finite, the distribution of compartment states for a 
particular population is a stochastic object, which at a given time $t$ depends 
on the state of all compartments present in the population. 
Calculating the probability distribution over the full population state would be 
exceedingly challenging due to the combinatorial explosion of states in such systems. 
Moreover, such high-dimensional object would be difficult to work with and analyze in practice. 
To address these challenges, we have recently proposed a more effective approach, 
which summarizes the high-dimensional population state in terms of a small number 
of population moments, such as the total number of compartments, or the total 
amount of a certain molecule across the population~\cite{Duso2020}. 
A key difference to the moment techniques mentioned above is that the resulting 
moments are stochastic, which is a consequence of the finite population size. 
To effectively access the statistical properties of the compartment population, 
we derived ordinary differential equations which capture the mean and variance 
of the stochastic population moments. 
As with conventional moment-based techniques, the resulting system of equations 
is not necessarily closed, which we have so far addressed using \emph{ad hoc} 
moment-closure approximations~\cite{Duso2020}. 

There are two main limitations of this approach. 
First, identifying suitable closure functions can be difficult in practice 
as they are typically found in a trial-and-error fashion. 
Second, it is currently restricted to systems with polynomial rate laws to ensure that 
the resulting moment equation hierarchy is self-contained
~\cite{schnoerr2017approximation,Duso2020}
. 
In the present work, we develop a systematic moment-approximation technique
which aims at addressing these two issues. 
The approach is based on a local polynomial approximation of the underlying rate-laws,
which for the considered class of systems factorizes into a product of 
two functions that depend on the compartment content and compartment 
number distribution, respectively. 
The resulting scheme always leads to a closed set of moment equations 
and is applicable also to systems involving non-polynomial rate-laws. 
We demonstrate our approach using several biologically-inspired models 
and analyze its accuracy and runtime. 
Our implementation of the moment-approximation technique is publicly 
available through a new release of
our previously developed symbolic moment generator \emph{Compartor}~\cite{pietzsch2021compartor}.

\section{Results}

\subsection{Stochastic compartment populations}\label{sec:stochcomp}

To describe the dynamics of a stochastic compartment population, we make use of our previously proposed formalism~\cite{Duso2020}. We define a compartment population as a 
collection of $N$ distinct entities, each one being associated with a
$d$-dimensional content variable $x \in \X \subseteq \N^d$.
Individual dimensions of $x$ typically correspond to the copy number of
a particular chemical species, but could encode also other discrete-valued features, such as compartment categories (e.g., cell types).

The state of the population can be represented by a \emph{number distribution}
function $n : \X \to \N$ that maps any possible value of the content variable
$x$ to the number of compartments $n(x)$  having that particular content.
The number distribution can be represented by a (typically infinite) 
multidimensional array $ \n = ( n(x) )_{x \in \X} $, where
the compartment number $n(x)$ corresponds to the element of $\n$ at index $x$,
i.e. $\n_x = n(x)$.

The compartment population can change dynamically according to a set of 
\emph{transition classes}.
Transition classes can encode changes in both the molecular content of a compartment
and in the population of compartments itself: for instance two molecules may react
to produce a third one within one compartment, or two compartments may fuse into
a single one, merging their contents.
Transition classes therefore generalise the concept of \emph{reactions} to compartmentalised systems. 

A transition class $c$ is defined by a set $x_c = \nolinebreak \{x^{(1)}_c, \ldots, x^{(r_c)}_c\}$ of 
reactant compartments, a set $y_c =  \nolinebreak \{y^{(1)}_c, \ldots, y^{(p_c)}_c\}$ of product
compartments and by a rate function $h_{c}(\n(t), x_c, y_c)$
which determines how likely a transition with a particular combination of reactant
and product compartments occurs per unit time. Throughout this work, we consider the compartment population to exhibit Markovian dynamics such that the rate functions depend only on the current state of the population $\n(t)$. 

Consider a compartment population whose dynamics follows a set of transition classes $\C$.
The time evolution of the number distribution can then be described as
\begin{equation}
	\n(t) = \n(0) + \sum_{c\in\C} \sum_{x_c, y_c} \Delta\n_c(x_c, y_c) R_{c, x_c, y_c}(t)
	\label{eq:populationChange}
\end{equation}
where $\n(0)$ is the initial state of the population at $t=0$ and $R_{c, x_c, y_c}(t)$ are 
counting processes with rate function $h_{c}(\n(t), x_c, y_c)$ tracking the number of transitions of type $c$ involving the particular sets of reactant- and product compartment $x_c$ and $y_c$ up to time $t$. The array $\Delta\n_c(x_c, y_c) = (-\sum_{x\in x_c} \delta_{x, s} + \sum_{y\in y_c} \delta_{y, s})_{s \in \X}$ captures how the population state $\n(t)$ changes when this particular transition happens. The process defined in (\ref{eq:populationChange}) can be understood as an instance of a measure-valued stochastic process \cite{dawson1993measure, bansaye2015stochastic}. Note that for compactness, we will drop the time-dependency of $\n(t)$ and all derived quantities in the following.

Throughout this work, we consider rate functions of the form
\begin{equation}\label{eq:ratefun}
	h_{c}(\n, x_c, y_c) = k_c \pi_c(y_c \mid x_c) g_c(x_c) \, w_{c, x_c}(\n) \;,
\end{equation}
with $k_c$ as a rate constant, $g_c$ as a function that depends exclusively 
on the content of the reactant compartments 
and $w_{c, x_c}:=w_c(n(x_c^{(1)}), \ldots, n(x_c^{(r_c)}))$ 
as a function that depends only on the population state $\n$. 
Throughout this work, we consider $w_{c, x_c}$ to be a multilinear polynomial, 
which arises combinatorially by counting all indistinguishable combinations 
of reactant compartments for a particular population state $\n$.
We refer to functions $w_{c, x_c}$ defined in this way as \emph{mass-action-like}.
The function $\pi_c$ denotes the \emph{outcome distribution} 
that determines how likely a particular set of product compartments is realized 
from a given set of reactant compartments. 
In the case of cell division, for examples, $\pi_c$ describes the partitioning of molecular content between daughter cells given the content of the mother cell right before division.
Concrete examples of transition classes 
and their rate functions will be provided later in our case studies.

We can observe from \cref{eq:ratefun} that the transition rate functions in 
the considered systems are \emph{separable}:
they factorise into a product of a function
that depends on the content of the involved compartments and
another function that depends on how many such compartments are available in the population.
We can therefore write
\begin{equation}
	h_c(\n, x_c, y_c) = f_c(x_c, y_c) \, w_{c,x_c}(\n)
\end{equation}
where $f_c(x_c, y_c)$ only depends on the content of the reactant and product 
compartments and
$w_{c,x_c}(\n)$ depends explicitly only on the number of compartments with the given contents ($x_c$), but not on the content itself. 

\subsection{Moment dynamics}\label{sec:momEq}
While \cref{eq:populationChange} describes the time evolution of the full
population state, this equation is difficult to deal with in practice. 
A more effective way to analyze compartmentalized systems is provided by moment
equations,
which capture certain statistical properties of the compartment population~\cite{ramkrishna2000population,Duso2020,lunz2022revisiting}. 
We define a \emph{population moment} of exponent $\gamma \, \in \, \N^d$ by
\begin{equation} \label{eq:momentDef}
	M^\gamma = \sum_{x} x^\gamma \, \n_x
	= \sum_{x} x_1^{\gamma_1} x_2^{\gamma_2} \ldots 
		x_d^{\gamma_d} \, \n_x \;.
\end{equation}
The \emph{order} of this moment is given by $|\gamma|$, which is the sum of all 
elements of the multi-index $\gamma$. Note that a population moment is a functional of the stochastic number distribution $\n$ and correspondingly, is itself stochastic~\cite{Duso2020}. 

A special population moment is the moment of order zero $M^{0}$, which counts the total number of compartments present in the system. In the following, we will refer to this moment as $N \coloneqq M^{0}$. Similarly, a moment of order one, i.e., $M^{e_j}$ with $e_j$ as the $j$-th unit vector, counts the total amount of the $j$-th species across all compartments.

Starting from \cref{eq:populationChange}, we can now derive a stochastic differential equation which captures the time-evolution of a population moment $M^\gamma$, 
\begin{equation}
	\de M^\gamma = \sum_{c\in\C} \sum_{x_c, y_c} \Delta M_c^\gamma(x_c, y_c) \de R_{c,x_c, y_c},
	\label{eq:stochMom}
\end{equation}
where $\Delta M_c^\gamma(x_c, y_c) = \sum_s s^{\gamma} \, \Delta n_{c, x_c, y_c}(s)$ denotes the instantaneous change of the moment $M^\gamma$ when the transition with compartment sets $x_c$ and $y_c$ from class $c$ happens. 
In practice, it is useful to study the expected behavior of a stochastic population moment.
To this end, we can take the expectation on both sides of \cref{eq:stochMom}, which yields
\begin{equation}
	\label{eq:baseDiffeq}
	\ddt{\esp{M^\gamma}} = \sum_c \besp{ \sum_{x_c, y_c} \Delta M_c^\gamma(x_c, y_c) \, h_c(\n, x_c, y_c) } \;.
\end{equation}
Eq. (\ref{eq:baseDiffeq}) is an ordinary differential equation that describes the time-evolution
of the \emph{expected moment} $\aesp{M^\gamma}$ and we refer to it as \emph{moment equation} throughout this work.

Using the separability of $h_c$ we can rewrite~\cref{eq:baseDiffeq} as
\begin{equation}
	\label{eq:separatedDiffeq_j}
	\ddt{\esp{M^\gamma}} = \sum_c 
		\besp{ \sum_{x_c,y_c} 
			\Delta M_c^\gamma(x_c,y_c)
			f_c(x_c,y_c) \, w_{c,x_c}(\n)
		} \;.
\end{equation}
Recalling~\cref{eq:ratefun}, we can further split the
function $f_c(x_c,y_c)$ into a product $\pi_c(y_c \mid x_c) \, g_c(x_c)$, 
where the rate function $g_c(x_c)$ only depends on the content of the 
reactant compartments, which is then multiplied by the probability 
$\pi_c$ of obtaining specific product compartments $y_c$. 
If we now carry out the summation over $y_c$, we obtain an expectation
over $\Delta M_c^\gamma$ conditionally on $x_c$. In particular, we obtain
\begin{align}
	\begin{split}
		\ddt{\esp{M^\gamma}} 
		= 
			\sum_c
			\besp{
				\sum_{x_c}
				\underbrace{
					\sum_{y_c} \Delta M_c^\gamma(x_c,y_c) \, \pi_c(y_c \mid x_c)
				}_{
					\E_{\pi_c} \left[ \Delta M_c^\gamma(x_c,y_c) \mid x_c \right]
				} \\
				\qquad \, g_c(x_c)
				\, w_{c,x_c}(\n)
			}
	\end{split} \notag \\
	\begin{split}
	= \sum_c
		\besp{ 
			\sum_{x_c}
			\underbrace{
				\E_{\pi_c} \left[ \Delta M_c^\gamma(x_c,y_c) \mid x_c \right]
			}_{
				\Delta M_c^\gamma(x_c)
			}
			\, g_c(x_c) \\
			\qquad \, w_{c,x_c}(\n)
		}
	\end{split} \notag \\
	\begin{split}
	= \sum_c
		\besp{ 
			\sum_{x_c}
			\Delta M_c^\gamma(x_c) \, g_c(x_c)
			\, w_{c,x_c}(\n)
		}
	\end{split} \label{eq:applicableDiffeq} \\
	\begin{split}
	= \sum_c
		\besp{ 
			\sum_{x_c}
			f_c^\gamma(x_c)
			\, w_{c,x_c}(\n)
		} \;,
	\end{split} \label{eq:separatedDiffeq}
\end{align}
showing that the individual summands are again separable into
two functions that depend
only on the compartment content
and the number distribution, respectively.
In the following, we refer to $f_c^\gamma$ and $w_{c,x_c}$ as \emph{content-} and \emph{state function}, respectively.
Moment equations for generic products of moments (e.g., $N^2$) exhibit 
the same separability, such that the following results that are based on the separability
property can be applied to them as well (\ref{supp:ito}).

\subsection{Challenges of deriving moment equations in practice}\label{sec:practicalChallenges}
When deriving moment equations for a particular system,
we can use~\cref{eq:separatedDiffeq}.
The two functions, $f_c^\gamma$ and $w_{c,x_c}$, can be
directly derived from the specifications of a given transition class $c$ and for any
given moment exponent $\gamma$.
However, difficulties may arise depending on their particular form.
These difficulties can be best explained by going through a series of representative examples.

\paragraph{Linear content- and state function.}
The simplest scenario occurs when both the content- and state functions are
at most linear polynomials.
In this case the resulting moment dynamics is always closed, 
similarly to what happens in moment equations for bulk chemical systems.
As an example, let us consider a simple degradation reaction occurring within the compartments
\begin{equation}
	[x] \react{\;h(\n;x)\;} [x-1] \;,
\end{equation}
with $h(\n;x)=x\,\n_x$ and assume that we are interested in analyzing the moments $N$ and $M^1$. The state function is given by
\begin{equation}
	w_{c,x_c} = \n_x
\end{equation}
and the content functions associated with $N$ and $M^1$ are
\begin{align}
	f_c^0(x) &= \Delta M_c^0 g_c(x) = 0 \cdot x \\
	\text{and}\quad f_c^1(x) &= \Delta M_c^1  g_c(x) = -1 \cdot x \;,
\end{align}
respectively. The moment equations for $\esp{N}$ and $\esp{M^1}$ are therefore given by
\begin{align}
	\ddt{\esp{N}} &= \aesp{\sum_x 0 \cdot x \cdot \n_x} = 0 \\
	\ddt{\esp{M^1}} &= \aesp{\sum_x -1 \cdot x \cdot \n_x} = - \aesp{M^1} \;,
\end{align}
which can be readily solved. This is because for linear content- and state functions, the moments appearing on the right-hand side of the equations 
cannot be of higher order than those on the left-hand side, ensuring that the resulting moment dynamics is closed.

\paragraph{Nonlinear content function.}
We next consider a more complex scenario, where the state function is still
linear, while the content function is a polynomial of higher degree.
As an example, we consider the case where compartments exit the system with a rate that
again linearly depends on their content:
\begin{equation}
	[x] \react{\;h(\n;x)\;} \emptyset \;,
\end{equation}
with $h(\n;x)=x\,\n_x$.
The relevant content- and state functions are now given by
\begin{align}
	f_c^0(x) &= \Delta M_c^0  g_c(x) = -1 \cdot x = -x \\
	f_c^1(x) &= \Delta M_c^1 g_c(x) = -x \cdot x = -x^2 \\
	w_{c,x_c}(\n) &= \n_x
\end{align}
and the equations for $\esp{N}$ and $\esp{M^1}$ become
\begin{align}
	\ddt{\esp{N}} &= \aesp{\sum_x -x \cdot \n_x} = -\aesp{M^1} \\
	\ddt{\esp{M^1}} &= \aesp{\sum_x -x^2 \cdot \n_x} = - \aesp{M^2} \;.
\end{align}
In this case, the equation for $\aesp{M^1}$ depends on the 
additional moment $\aesp{M^2}$ because the content function
$f_c^1(x)$ is a second degree polynomial. 
Deriving the equation for $\aesp{M^2}$ yields
\begin{align}
	\ddt{\esp{M^2}} &= \aesp{\sum_x -x^2 \cdot x \cdot \n_x} = - \aesp{M^3} \;,
\end{align}
which in turn depends on the additional moment $\aesp{M^3}$. More generally, it turns out that the equation for any $\aesp{M^k}$ depends on $\aesp{M^{k+1}}$, leading to an infinite-dimensional system of moment equations.
This phenomenon is generally known as
\emph{closure problem}~\cite{schnoerr2017approximation,ramkrishna2000population}.

\paragraph{Nonlinear state function.}
We now consider a scenario where the state function is nonlinear. This is for instance the case when two compartments fuse, i.e., 
\begin{equation}
	[x] + [x'] \react{\;h(\n;x,x')\;} [x+x'] \;.
\end{equation}
In this case, the state function is
\begin{equation}
	w_{(x,x')}(\n) = \frac{\n_x (\n_{x'} - \delta_{x,x'}) }{ (1 + \delta_{x,x'}) } \;,
\end{equation}
which counts the number of all possible interactions of two reactant compartments
over unique combinations of $x, x'$.
Let us also assume that the fusion process is independent of the reactants' content.
Therefore we have $h(\n; x, x') = k \; w_{(x,x')}(\n) \;$.

Since the propensity of this transition class is independent of the content of 
the two reactant compartments, the content functions are 
$f_c^0 = -1 \cdot k$ and $f_c^1 = 0 \cdot k \;$.

The moment equations for $\esp{N}$ and $\esp{M^1}$ are therefore given by
\begin{align}
	\begin{split}
		\ddt{\esp{N}} 
		&= \aesp{\sum_{\{x,x'\}} -1 \cdot k \cdot 
			\frac{\n_x (\n_{x'} - \delta_{x,x'})}{(1 + \delta_{x,x'})} } \\
		&= \aesp{\sum_{x,x'} -1 \cdot k \cdot \frac{\n_x (\n_{x'} - \delta_{x,x'})}{2} } \\
		&= \aesp{\sum_{x,x'} -\frac{k}{2} \cdot (\n_x\n_{x'} - \delta_{x,x'}\n_x) } \\
		&= -\frac{k}{2}\aesp{N N} +\frac{k}{2}\aesp{N}	 
	\end{split}\label{eq:nonlinStateFunProblem_N} \\
	\begin{split}
		\ddt{\esp{M^1}} 
		&= \aesp{\sum_{x,x'} 0 \cdot k \cdot \frac{\n_x (\n_{x'} - \delta_{x,x'})}{2} } = 0 \;.
	\end{split}\label{eq:nonlinStateFunProblem_M1}
\end{align}
Notice that in the first line of~\cref{eq:nonlinStateFunProblem_N} the sum is taken
over unique combinations of $x$ and $x'$ (referred to as $\{ x, x' \}$), while in the remaining lines of~\cref{eq:nonlinStateFunProblem_N} and in~\cref{eq:nonlinStateFunProblem_M1} the sum is rewritten to go over all combinations of $x$ and $x'$, which requires the state function to be corrected for double-counting.

Notice that~\cref{eq:nonlinStateFunProblem_N} for $\aesp{N}$ depends on
$\n_x\n_{x'}$, 
making the state function a second-degree polynomial.
This causes $\aesp{N}$ to depend on the moment 
$\aesp{(N)^2} = \aesp{NN}$. 
An equation for $\aesp{NN}$ can be derived using It\^o's rule for counting processes 
(see \ref{supp:ito}):
\begin{align}
	\begin{split}
		\ddt{\esp{NN}} 
		&= \aesp{\sum_{x,x'} \left[ (N-1)^2 - (N)^2 \right] \cdot k 
			\cdot \frac{\n_x (\n_{x'} - \delta_{x,x'})}{2} } \\
		&= \aesp{\sum_{x,x'} \left[ -2N + 1 \right] \cdot \frac{k}{2} 
			\cdot (\n_x\n_{x'} - \delta_{x,x'}\n_x) } \\
		&= -k\aesp{NNN} + \frac{3}{2}\aesp{NN} - \frac{1}{2}\aesp{N} \;.
	\end{split}
\end{align}
This demonstrates that in this scenario the equation for any $\aesp{(N)^k}$ 
depends on $\aesp{(N)^{k+1}}$,
leading again to an infinite set of moment equations.
Note however that this closure problem is different from the one of the previous
scenario as it originates from \emph{products} of moments (e.g. $NNN$), 
as opposed to higher order moments such as $M^k$. This illustrates that the compartmentalised systems considered here display two types of 
closure problems, which is an important difference to bulk chemical systems
or compartmentalized systems in the large $N$ limit.

We remark that for general polynomial content- and state functions, both closure problems can occur simultaneously, leading to products of multiple higher-order moments such as $\aesp{NM^3}$ or $\aesp{M^{2,0}M^{0,3}}$, for instance. 

\paragraph{Non-polynomial functions.}
In many practical cases, the involved rate functions may not be of polynomial form. This is the case, for instance, when considering enzyme kinetics inside compartments (e.g., Michaelis-Menten or Hill functions) or compartment fusion driven by non-polynomial coagulation kernels. 
In this case, the functions $f_c^\gamma$ and/or $w_{c,x_c}$ take more general, non-polynomial forms, 
causing the moment dynamics to depend on terms which cannot be expressed as moments at all. 
In order to study such systems using moment equations, suitable polynomial approximations are necessary.

\subsection{A local polynomial (moment) approximation for compartmentalised reaction systems (LPAC)} \label{sec:approxStrategy}
The demonstration of the closure-related challenges 
outlined in~\cref{sec:practicalChallenges} also suggests a possible remedy.
In case of polynomial content- and state functions  $f_c^{\gamma}$ and $w_{c, x_c}$, their degrees dictate the
order of the moments and moment products appearing on the right-hand side of (\ref{eq:separatedDiffeq}). 
Therefore, constraining the degree of $f_c^{\gamma}$ and $w_{c, x_c}$ will bound
 both the order of the moments and the degree of moment products. 
If we apply the same constraints consistently to all the moments involved, this will ultimately lead to a closed system of moment equations.

To demonstrate this idea, consider maximum degrees $\delta_c$  and $\delta_s$ for the content- and state function, respectively. We can first derive equations for a given set of moments using~\cref{eq:applicableDiffeq} and put them into their separated form. The resulting content- and state functions are then approximated by polynomials of degree
$\delta_c$ and $\delta_s$. Taking the expectations will lead to a linear combination of expected moments. These may involve additional expected moments, for which new equations have to be derived. The content- and state functions appearing on the right-hand side of these new equations are approximated again using polynomials of order $\delta_c$ and $\delta_s$. This procedure is repeated until no new expected moments appear, resulting in a closed system of equations.

The simplest way to obtain polynomial approximations of the functions 
$f_c^{\gamma}$ and $w_{c, x_c}$ is to use truncated Taylor series around suitable expansion points $\hat{x}$ and $\hat{\n}$. 
To derive explicit expressions for the approximate moment equations, we first define the following shorthand notation
\begin{align}
	\D_{k,i}^{\alpha} \coloneqq{}& \frac{\partial^{\alpha_{k,i}}}{\partial (x_c)_{k,i}} \\
	\xi_{k,i} \coloneqq{}& (x_c)_{k,i} - \hat{x}_{i} \label{eq:defXi} \\
	\bmD_{k}^{\beta} \coloneqq{}& \frac{\partial^{\beta_k}}{\partial \n_{(x_c)_k}} \\
	\bmeta_{k} \coloneqq{}& \n_{(x_c)_k} - \hat{\n}_{(x_c)_k} \;, \label{eq:defEta}
\end{align}
where $x_c$ is the set of reactant compartments of size $r_c$ corresponding to the $c$-th transition class and $(x_c)_{k, i}$ denotes the $i$-th chemical species of the $k$-th compartment in $x_c$.
The exponents $\alpha_{k,i}$ and $\beta_k$ denote the differentiation orders
associated with the content- and state functions, respectively.

The Taylor expansions of the content- and state functions can then be written as
\begin{align}
	f_c^{\gamma}(x_c) ={}& 
	\sum_{\alpha_{1,1}=0}^\infty \ldots \sum_{\alpha_{r_c,1}=0}^\infty \cdots 
	\sum_{\alpha_{1,d}=0}^\infty \ldots \sum_{\alpha_{r_c,d}=0}^\infty \bigg[ \notag \\
	{}&(\D_{1,1}^{\alpha} \cdots \D_{r_c,d}^{\alpha}) 
	\left. f_c^{\gamma} \right|_{(\hat{x}, \ldots, \hat{x})} \notag \\
	 {}& 
		\frac{
			\left( \xi_{1,1}\right)^{\alpha_{1,1}}
			\cdots \left( \xi_{r_c,d}\right)^{\alpha_{r_c,d}}
		}{
			\alpha_{1,1}! \cdots \alpha_{r_c,d}!
		} \bigg] \label{eq:taylorContent} \\
	={}& \sum_{\alpha} \D^{\alpha} 
		\left. f_c^{\gamma} \right|_{\hat{x}} 
		\, \frac{\xi^{\alpha}}{\alpha!} \label{eq:taylorContentCompact}
\end{align}
and
\begin{align}
	w_{c,x_c}(\n) ={}&
	\sum_{\beta_1=0}^\infty \ldots \sum_{\beta_{r_c}=0}^\infty \bigg[ \notag \\
	{}&(\bmD_{1}^{\beta} \cdots \bmD_{r_c}^{\beta}) 
	\left. w_{c,x_c} \right|_{(\hat{\n}_{(x_c)_1}, \ldots, \hat{\n}_{(x_c)_{r_c}} )} \notag \\
	 {}& 
		\frac{
			\left( \bmeta_1\right)^{\beta_1}
			\cdots \left( \bmeta_{r_c}\right)^{\beta_{r_c}}
		}{
			\beta_1! \cdots \beta_{r_c}!
		} \bigg] \label{eq:taylorState} \\
	={}& \sum_{\beta} \bmD^{\beta} 
		\left. w_{c,x_c} \right|_{\hat{ \n }} 
		\, \frac{\bmeta^{\beta}}{\beta!} \;, \label{eq:taylorStateCompact}
\end{align}
respectively.

If we substitute the expansions, \cref{eq:taylorContentCompact,eq:taylorStateCompact},
into the separated moment equation, \cref{eq:separatedDiffeq}, we finally obtain
the full explicit form of the Taylor-based moment approximation
\begin{equation}
\begin{split}
	\ddt{\esp{M^\gamma}} 
	={}& \sum_c
		\besp{ 
			\sum_{x_c}
			f_c^\gamma(x_c)
			\, w_{c,x_c}(\n)
		} \\
	\approx{}& \sum_c \sum_{\alpha,\beta}\;
		\besp{ 
			\sum_{x_c} \\
	{}&			\D^{\alpha} 
				\left. f_c^{\gamma} \right|_{\hat{x}} 
				\, \bmD^{\beta} 
				\left. w_{c,x_c} \right|_{\hat{ \n }} 
			\, \frac{\xi^{\alpha}}{\alpha!}
				\, 
				\frac{\bmeta^{\beta}}{\beta!}
		} \;.
\end{split} \label{eq:finalTaylorApprox}
\end{equation}

We now choose the first expasion point as $\hat{x} = (\hat{x}_i)_i$ such that its components are
\begin{equation}
	\hat{x}_i = \frac{ \esp{M^{\e_i}} }{ \esp{N} } \;,
\end{equation}
where $\e_i$ is the $i$-th unit vector. 
This can be understood as an estimate of the mean compartment content.
For the second expansion point we choose the expected number distribution
\begin{equation}
	\hat{\n} = \esp{\n} \;.
\end{equation}
This is particularly important as it allows us to conveniently translate the
$\sum_x x^\gamma \, \bmeta$ terms arising in the Taylor expansion~\eqref{eq:finalTaylorApprox} 
into functions of moments as
\begin{equation}
	\begin{split}
		\sum_x x^\gamma \, \bmeta 
		&= \sum_x x^\gamma \, \left( \n_x - \esp{\n_x} \right) \\
		&= \sum_x x^\gamma \, \n_x - \sum_x x^\gamma \, \esp{\n_x} \\
		&= \sum_x x^\gamma \, \n_x - \aesp{ \sum_x x^\gamma \, \n_x } \\
		&= M^\gamma - \aesp{ M^\gamma } \;.
	\end{split}
\end{equation}

We can now plug the chosen expansion points into~\eqref{eq:finalTaylorApprox}
to obtain the final compact form:
\begin{align}
	\ddt{\esp{M^\gamma}}
	=& \sum_c \sum_{\alpha} \sum_{\beta \leq \bm{e}}\;
		\sum_{k = \bm{0}}^{\alpha} \; \notag \\
	{}&	\D^{\alpha} 
				\left. f_c^{\gamma} \right|_{\hat{x}} 
				\, \bmD^{\beta} 
				\left. w_{c,x_c} \right|_{\esp{ \n }} 
			\, \binom{\alpha}{k}
				\left( -\hat{x} \right)^{k} \notag \\
	{}&	\aesp{ \left( M^{\alpha-k} - \aesp{M^{\alpha-k}} \right)^{\beta}
		} \;, \label{eq:momentTaylorApprox}
\end{align}
where we used the standard multi-index notation for
all operators involving tuples to arrive at the final compact form.
See \ref{supp:multi-index} for the precise definition of
all multi-index operations used in our work.
Note that~\cref{eq:momentTaylorApprox} relies on the assumption that the state
function $w_{c,x_c}$ is multilinear, as indicated by 
the summation domain $\beta \leq \e$, i.e. $\beta_i \leq 1$ for all $i$.
If we now truncate~\cref{eq:momentTaylorApprox} 
such that $|\alpha| \leq \delta_c$ and $|\beta| \leq \delta_s$ we 
obtain the final approximated moment equation for $\aesp{M^\gamma}$.

\section{Case studies} \label{sec:caseStudies}
To test the proposed moment expansion method, we applied it to three different case studies. These examples are chosen to resemble characteristic features of biological systems such as non-linear propensities, interactions among compartments and feedback loops. We analyze the accuracy of our moment-approximation approach, by comparing it to exact Monte Carlo simulation.
Moment equations were derived automatically by the 
\emph{Compartor}\cite{pietzsch2021compartor} toolbox, which was extended to support 
the presented LPAC moment-expansion method. 
The resulting equations were solved using the \emph{DifferentialEquations.jl}\cite{rackauckas2017differentialequations}
 package of the Julia programming language\cite{Julia2017}.

\subsection{Binary birth-death-fusion process} \label{sec:bbdc}
\begin{figure}[t]
	\centering
	\includegraphics[scale=\figscale]{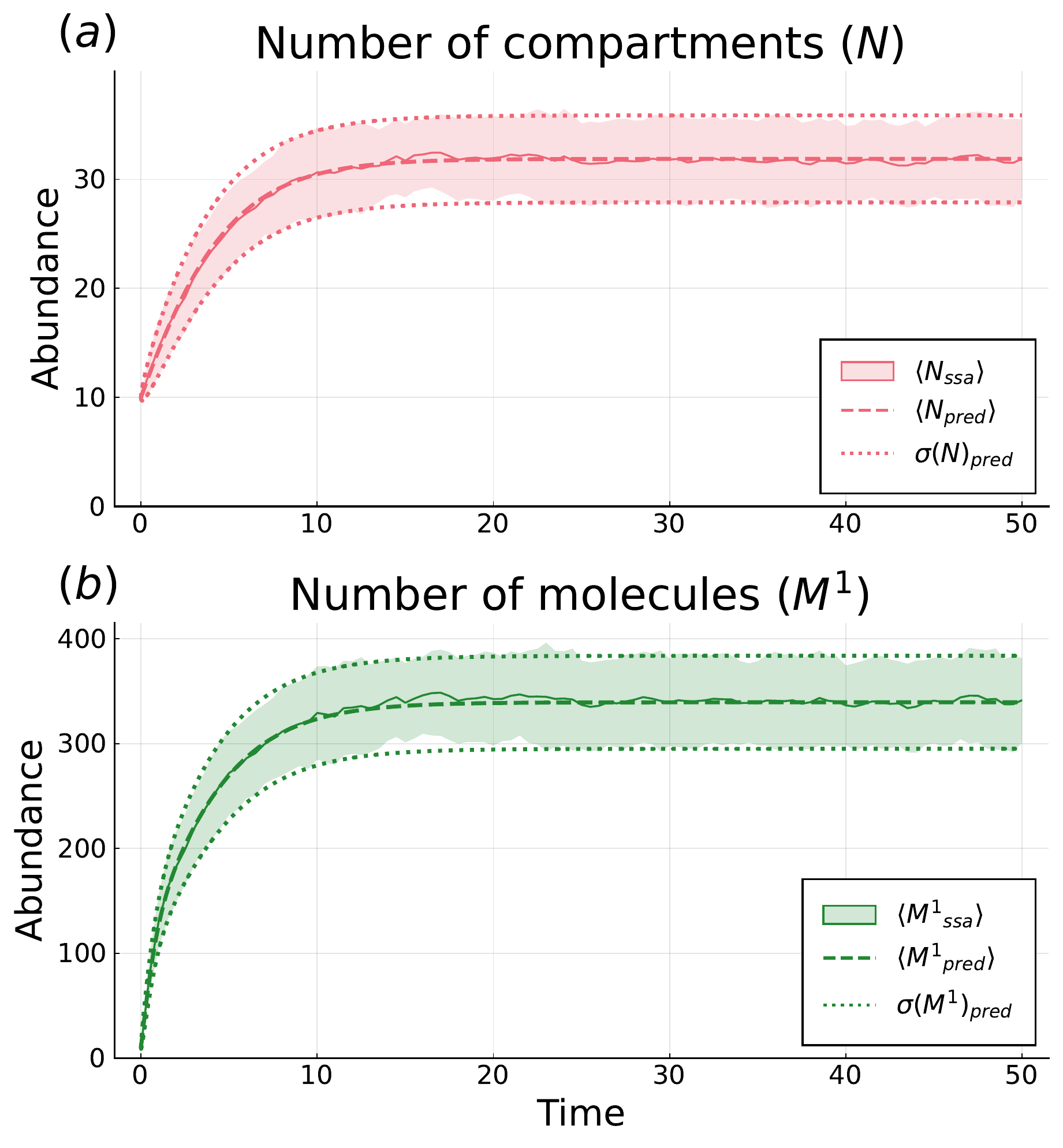}%
	\caption{Comparison of the results obtained by averaging 100 trajectories of the 
	stochastic simulation algorithm (SSA), shown by the solid line and shaded area,
	with the predictions of our moment-expansion method, shown by the dashed line
	and dotted boundaries. The solid and dashed lines denote average values, while
	the shaded and dot-bordered areas are the regions within one standard deviation.
	Panel (\emph{a}) shows the statistics of the number of compartments ($N$)
	and panel (\emph{b}) shows the statistics of the total number of molecules in the system ($M^1$).}%
	\label{fig:bbdc}
\end{figure}
\begin{table}
	\begin{tabularx}{\linewidth}{Sl Y Y} 
		Case study & Statistic & Error \\
		\hline

		\arrayrulecolor{gray!50} 

		\multirow{2}{*}{bBDF} & $N$ & $6\,\%$ \\
		 & $M^1$ & $7\,\%$ \\
		\hline

		\multirow{3}{*}{sAIC} & $N$ & $30\,\%$ \\
		 & $Z_1$ & $34\,\%$ \\
		 & $Q$ & $30\,\%$ \\
		\hline

		\multirow{4}{*}{MR} & $N$ & $7\,\%$ \\
		 & $M^{1,0}$ & $49\,\%$ \\
		 & $M^{1,1}$ & $65\,\%$ \\

		\arrayrulecolor{black} 
	\end{tabularx}
	\caption{
		Error quantifications in the three case studies: 
		\emph{Binary birth-death fusion} (bBDF),
		\emph{Shared Antithetic Integral Controller} (sAIC) and
		\emph{Mutual gene repression} (MR).
		Errors were calculated by taking the absolute distance between the moments
		estimated from Monte Carlo simulations and LPAC, dividing it by the Monte Carlo 
		standard deviation and averaging the resulting value across all time points.
	}
	\label{tab:errors}
\end{table}

We first tested our method on a simple system that features two nonlinear
propensity functions. Compartments contain a single chemical species $\mathrm
{X}$ which is synthesized with a constant rate $k_b$. For molecular turnover,
we consider an annihilation reaction $2\mathrm{X}\rightarrow \emptyset$ with
rate constant $k_d$. Compartments enter the system at a rate $k_I$ and their
initial content is distributed according to an intake distribution $\pi_I$,
which we consider to be a Poisson distribution with mean $\lambda$. Moreover,
two compartments can fuse with each other with rate constant $k_F$. In total,
the system is given by the transition classes
\begin{align}
	\emptyset &\react{h_I(\n; y)} [y] && \text{(Intake)} \notag \\
	[x] + [x'] &\react{h_F(\n; x,x')} [x + x'] && \text{(Fusion)} \notag \\
	[x] &\react{h_b(\n; x)} [x+1] && \text{(Birth)} \notag \\
	[x] &\react{h_d(\n; x)} [x-2] \;, && \text{(Death)} \notag
\end{align}
with propensity functions
\begin{align}
	h_I(\n; y) &= k_I \, \pi_I(y) \label{eq:bbdc:hI} \\
	h_F(\n; x,x') &= k_F \, \frac{\n_x \, (\n_{x'} - \delta_{x,x'})}{1 + \delta_{x,x'}} \label{eq:bbdc:hE} \\
	h_b(\n; x) &= k_b \, \n_x \label{eq:bbdc:hb} \\
	h_d(\n; x) &= k_d \, \frac{x \, (x-1)}{2} \, \n_x \;, \label{eq:bbdc:hd} 
\end{align}
where $x$, $x'$ and $y$ are variables representing the compartments' 
chemical content.
Here we observe that both the fusion- and annihilation transitions involve propensities that
are nonlinear in state and content, respectively.

Our goal is to use our moment-expansion method to study the mean and variance of the population moments $N$ and $M^1$. 
To this end, we require equations for the averages 
$\aesp{N}$ and $\aesp{M^1}$ as well as the second-order moments 
$\aesp{N^2}$ and $\aesp{(M^1)^2}$ from which variances can 
be obtained subsequently.
However, deriving the equations for these moments leads to a closure
problem due to the nonlinear propensities.
We therefore employ a second-order polynomial 
approximation based on the Taylor series, which leads to a closed system of equations for the moments
$\aesp{N}$, $\aesp{N^2}$, $\aesp{M^1}$, $\aesp{(M^1)^2}$, $\aesp{M^2}$, $\aesp{(M^2)^2}$, $\aesp{NM^1}$, $\aesp{NM^2}$ and $\aesp{M^1M^2}$.
For further details refer to \ref{supp:bbdc}.

\Cref{fig:bbdc} shows that the approximate statistics of $N$ and $M^1$ 
(dashed and dotted lines) are in close agreement with Monte Carlo simulations 
(solid lines and shaded areas). 
Quantitative error measures are reported in~\cref{tab:errors}. 
On our installation, generating $n=100$ Monte Carlo samples took $4.7$s while solving the moment equations 
took $6.6 \cdot 10^{-4}$s, corresponding to a speedup of about 7121 times.

\subsection{Shared Antithetic Integral Controller} \label{sec:saic}
\begin{figure}[t]
	\centering
	\includegraphics[scale=\figscale]{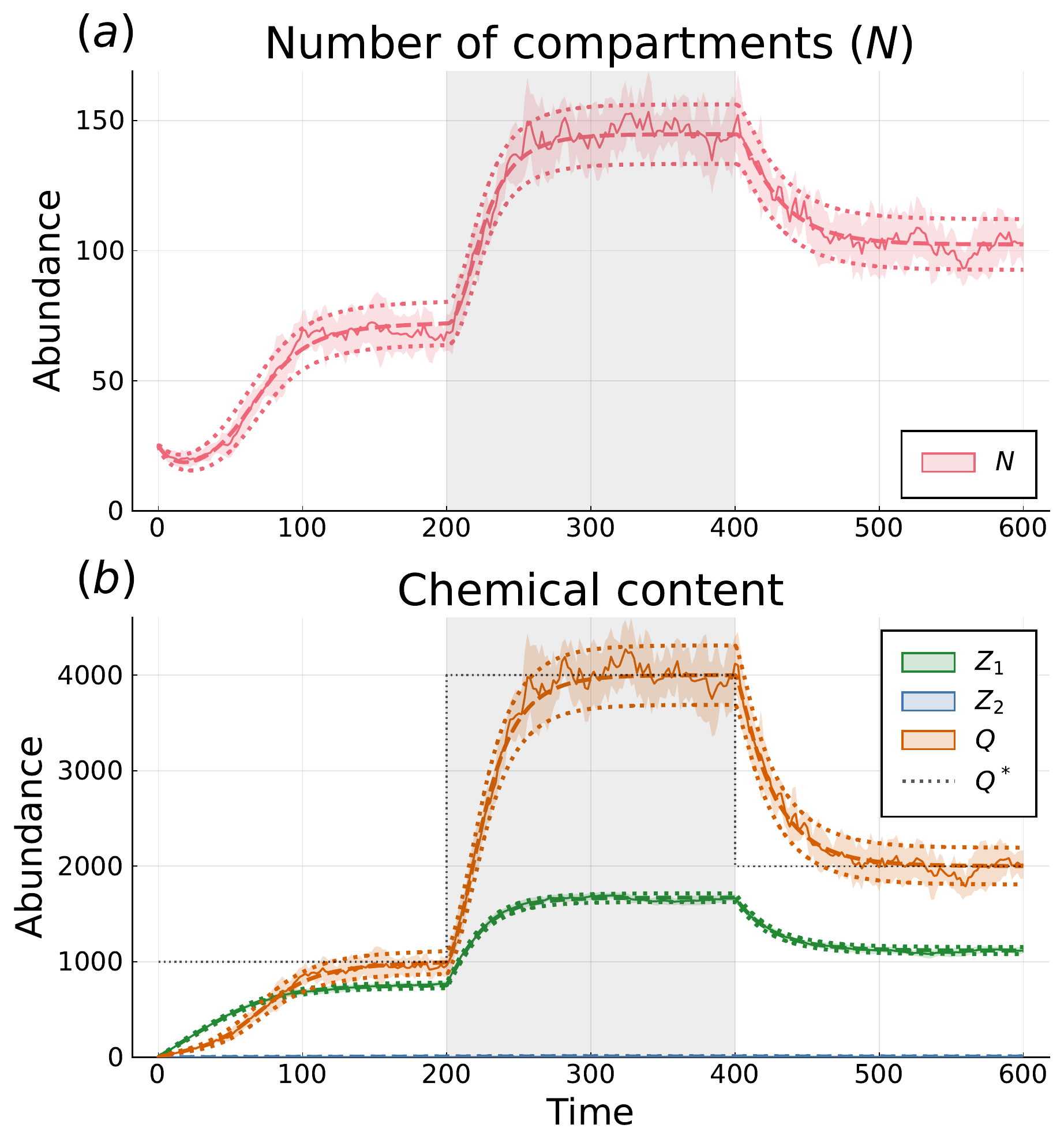}%
	\caption{Comparison of the results obtained by averaging 8 SSA trajectories, 
	shown by the solid line and shaded area,
	with the predictions of our moment-expansion method, shown by the dashed line
	and dotted boundaries. The solid and dashed lines denote average values, while
	the shaded and dot-bordered areas are the regions within one standard deviation.
	Panel (\emph{a}) shows the statistics of the number of compartments ($N$)
	and panel (\emph{b}) shows the statistics of the total number of
	molecules of different species in the system %
	($Z_1$, $Z_2$ and $Q$), %
	together with the controller's setpoint $Q^*$ indicated by the grey dotted line.
	}%
	\label{fig:SAIC}
\end{figure}

As a second case study we apply our approximation approach to a population-level feedback control motif that we analyzed previously \cite{duso2021shared}. This system is based on the \emph{antithetic integral control} motif -- a general biochemical control strategy that ensures robust perfect adaptation \cite{briat2016antithetic}. In contrast to the original scheme, the considered 
control system --- termed \emph{shared antithetic integral controller (sAIC)} --- acts at the multicellular level and thereby provides a means to robustly regulate population-level features such as the total amount of a produced species or the number of cells in the population. In our original work \cite{duso2021shared} we applied the sAIC to a simple cell population model and we adopt a similar model for the present case study.
\enlargethispage{\baselineskip} 

Here we consider a population of cells undergoing division and apoptosis and
within these cells a compartmentalised species $Q$ is produced and degraded.
This system is extended by the sAIC circuit by introducing
two bulk controller species $Z_1$, $Z_2$ and four additional transition
classes.
In total, the considered system is given by the following transition network
\begin{align}
	[x] &\react{h_D(\n; x,y)} [y] + [x-y] & \text{(Division)} \notag \\
	[x] + [x'] &\react{h_A(\n; x)} [x] & \text{(Apoptosis)} \notag \\
	[x] &\react{h_p(\n; x)} [x + 1] & \text{(Production)} \notag \\
	[x] &\react{h_d(\n; x)} [x - 1] & \text{(Degradation)} \notag \\
	\emptyset &\react{h_{ref}(\n; x)} Z_1 & \text{(Reference)} \notag \\
	Z_1 + [x] &\react{h_{act}(\n; x)} Z_1 + [x + 1] & \text{(Actuation)} \notag \\
	[x] &\react{h_{meas}(\n; x)} [x] + Z_2 & \text{(Measurement)} \notag \\
	Z_1 + Z_2 &\react{h_{comp}(\n; x)} \emptyset & \text{(Comparison)} \notag
\end{align}
with propensity functions
\begin{align}
	h_D(\n; x,y) &= k_D \, \pi_D(y \mid x) \, x \, \n_x \label{eq:saic:hD} \\
	h_A(\n; x,x') &= k_A \, \frac{\n_x (\n_{x'} - \delta_{x,x'})}{1 + \delta_{x,x'}} \label{eq:saic:hA} \\
	h_p(\n; x) &= k_p \, \n_x \label{eq:saic:hp} \\
	h_d(\n; x) &= k_d \, x \, \n_x \label{eq:saic:hd} \\
	h_{ref}(\cdot) &= k_{ref} \label{eq:saic:hRef} \\
	h_{act}(\n, Z_1; x) &= k_{act} \, Z_1 \, \n_x. \label{eq:saic:hAct} \\
	h_{meas}(\n; x) &= k_{meas} \, x \, \n_x \label{eq:saic:hMeas} \\
	h_{comp}(Z_1, Z_2) &= k_{comp} \, Z_1 Z_2 \;, \label{eq:saic:hComp}
\end{align}
where $x$, $x'$ and $y$ are variables representing the compartments' 
chemical content, $k_D$, $k_A$, $k_p$, $k_d$, 
$k_{ref}$, $k_{act}$, $k_{meas}$ and $k_{comp}$ are the rate constants
for the respective transition classes and $\pi_D$ is the outcome distribution for
cell division, 
where content is uniformly distributed among daughter cells.
The \emph{Division} and \emph{Apoptosis} transition classes capture
the dynamics of the cell population, where individual cells can divide and turn over.
\emph{Production} and \emph{Degradation} describe the dynamics of the
cell-internal species $Q$.
Feedback-control is achieved via the four additional transition classes:
\emph{Reference} and \emph{Actuation} are responsible for setting the target
level of $Q$ and promoting its production,
while \emph{Measurement} and \emph{Comparison} provide a way to sense the current 
value of $Q$ and downregulate its production. As we have shown previously, this 
circuit regulates the average total number of molecules $\aesp{Q}$, i.e. $\aesp{M^1}$,
to tunable set points $Q^* = k_{ref} / k_{meas}$, irrespective of all other system parameters.
For further details, the reader may refer to our original work \cite{duso2021shared} and \ref{supp:SAIC}. 

\Cref{fig:SAIC} shows a controller with setpoints $Q^*$ that change at time $t = 200$
and $t = 400$, and the total mass of the controlled species $Q$ follows accordingly.
Moreover, the statistics of the cell number and the total number 
of molecules are again approximated accurately by the moment-expansion method. 
Quantitative error estimates can be found in~\cref{tab:errors}.

\subsection{Mutually repressing gene circuit in a cell population} \label{sec:MutualRepression}
\begin{figure*}[t] 
	\centering
	\centerline{
		\includegraphics[scale=\widefigscale]{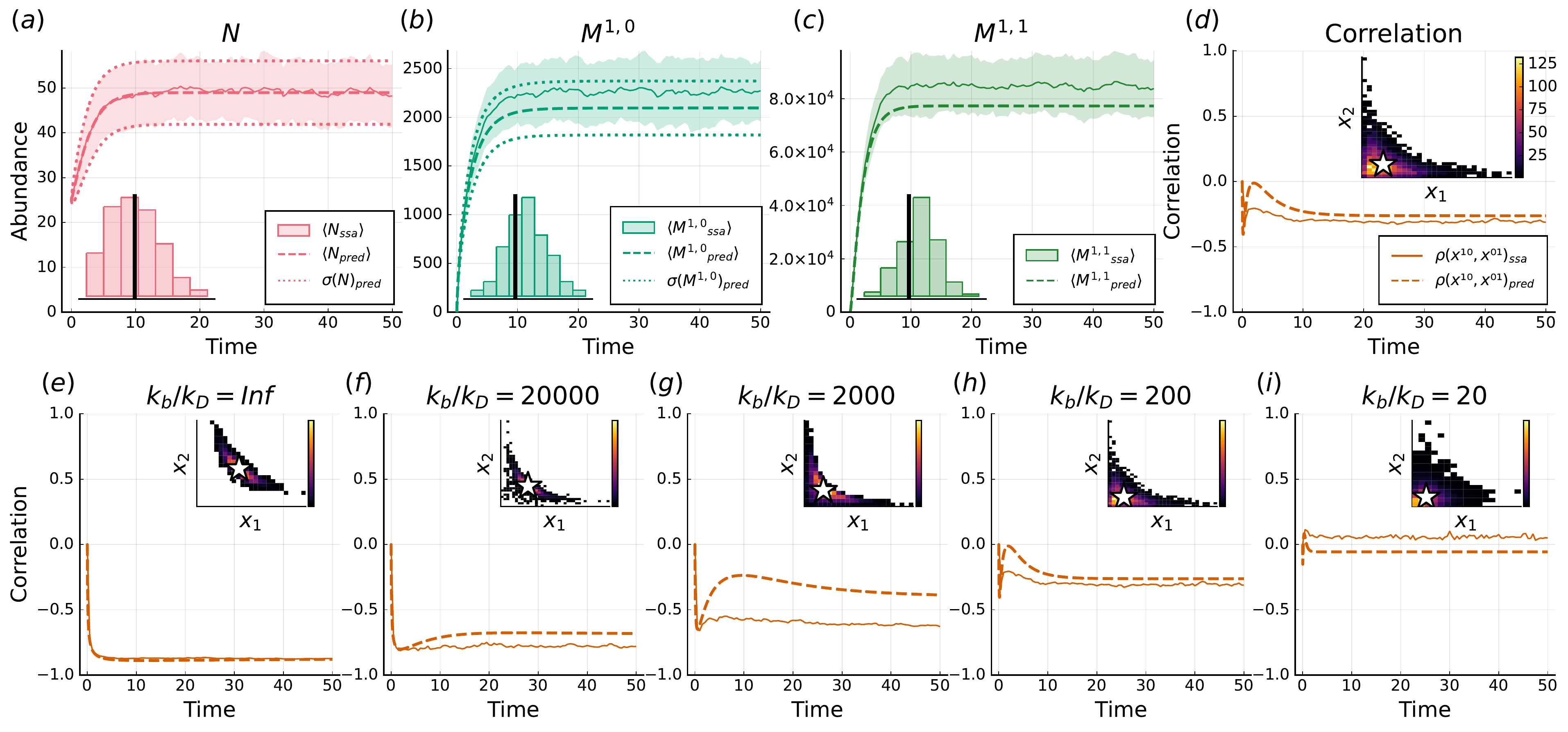}}%
	\caption{
	Comparison of the results obtained by averaging 128 SSA trajectories, 
	shown by the solid line and shaded area,
	with the predictions of our moment-expansion method, shown by the dashed line
	and dotted boundaries. The solid and dashed lines denote average values, while
	the shaded and dot-bordered areas are the regions within one standard deviation.
	Panels (\emph{a-d}) show the statistics of
	(\emph{a}) the number of compartments $N$,
	(\emph{b}) the total abundance of the first chemical species $M^{1,0}$, 
	(\emph{c}) the $M^{1,1}$ moment and
	(\emph{d}) the Pearson correlation coefficient for the two chemical species.
	Insets of panels (\emph{a,b,c}) respectively show the histograms of the $N$, $M^{1,0}$ 
	and $M^{1,1}$ moments across the stochastic trajectories, with the black solid
	line indicating the predicted mean value.
	The inset of panel (\emph{d}) shows the 2D histogram of the chemical content of compartments
	from all trajectories
	at the final timepoint of the SSA simulation, with the star symbol indicating the
	Taylor expansion point $(\sfrac{ \esp{M^{1,0}} }{ \esp{N} }, \sfrac{ \esp{M^{0,1}} }{ \esp{N} })$.
	Panels (\emph{e-i}) show the Pearson correlation and final-timepoint 2D histogram
	of the system in different $k_b/k_D$ regimes: 
	from (\emph{e}) no cell turnover occurring with $k_b/k_D = \infty$, 
	increasing the cell turnover speed (\emph{f-h}) to the fastest (\emph{i}) with $k_b/k_D = 20$.
	}
	\label{fig:MutualRepression}
\end{figure*}

In our last case study, we want to test our approach for a system that contains multiple chemical species that exhibit complex correlations.
To this end, we take inspiration from developmental biology and consider a population of dividing cells that express two mutually repressing genes.
Cell turnover is considered to be contact-dependent, where cells undergo apoptosis upon interacting with another cell from the population. 
The overall system is given by the transition classes
\begin{align}
	[x] &\react{h_D(\n; x,y)} [y] + [x-y] & \text{(Division)} \notag \\
	[x] + [x'] &\react{h_A(\n; x)} [x] & \text{(Apoptosis)} \notag \\
	[x] &\react{h_{p_1}(\n; x)} [x + (1,0)] & \text{(Production 1)} \notag \\
	[x] &\react{h_{d_1}(\n; x)} [x - (1,0)] & \text{(Degradation 1)} \notag \\
	[x] &\react{h_{p_2}(\n; x)} [x + (0,1)] & \text{(Production 2)} \notag \\
	[x] &\react{h_{d_2}(\n; x)} [x - (0,1)] & \text{(Degradation 2)} \notag
\end{align}
with propensity functions
\begin{align}
	h_D(\n; x,y) &= k_D \, \pi_D(y \mid x) \, \n_x \label{eq:MR:hD} \\
	h_A(\n; x,x') &= k_A \, \frac{\n_x (\n_{x'} - \delta_{x,x'})}{1 + \delta_{x,x'}} \label{eq:MR:hA} \\
	h_{p_1}(\n; x) &= k_p \, \frac{k_{R_1}}{k_{R_1} + x_2} \, \n_x \label{eq:MR:hp1} \\
	h_{d_1}(\n; x) &= k_d \, x_1 \, \n_x \label{eq:MR:hd1} \\
 	h_{p_2}(\n; x) &= k_p \, \frac{k_{R_2}}{k_{R_2} + x_1} \, \n_x \label{eq:MR:hp2} \\
	h_{d_2}(\n; x) &= k_d \, x_2 \, \n_x \;, \label{eq:MR:hd2} 
\end{align}
where $x$, $x'$ and $y$ are variables representing the compartments' chemical content, 
$k_D$, $k_A$, $k_p$ and $k_d$ are the rate constants for the respective transition classes, 
$k_{R_1}$ and $k_{R_2}$ are Michaelis-Menten constants and 
$\pi_D$ is the outcome distribution for cell division, 
where content is uniformly distributed among daughter cells.

Since the two internal species mutually repress each other (eq. (\ref{eq:MR:hp1}) and (\ref{eq:MR:hp2})), we generally expect negative correlations between them. Here we want to study how the cell population 
dynamics impacts this correlation and to what extent our moment-approximation method is able to capture this. 
As a measure for correlation, we use the Pearson correlation coefficient, 
which can be estimated from the approximate population moments as
\begin{align}
	\rho(x_1,x_2) &{}= \frac{ \cov(x_1,x_2) }{ \sqrt{\var(x_1)} \, \sqrt{\var(x_2)} } \notag \\
	&{}\approx \frac{ \aesp{M^{1,1}} - \frac{ \aesp{M^{1,0}} \, \aesp{M^{0,1}} }{ \aesp{N} } }%
	{ \sqrt{ \aesp{M^{2,0}} - \frac{\aesp{M^{1,0}}^2}{\aesp{N}} } \, \sqrt{ \aesp{M^{0,2}} - \frac{\aesp{M^{0,1}}^2}{\aesp{N}} } }
\end{align}
(for details see \ref{supp:correlation}).

\Cref{fig:MutualRepression}, panels (\emph{a-d}), 
shows the dynamics of the system in a regime where cell turnover 
and gene expression take place with a timescale ratio of $k_b/k_D = 200$, 
which is roughly the magnitude at which many physiological
processes take place~\cite{shamir2016characteristic}.
The respective quantitative error estimates can be found in~\cref{tab:errors}. 
Panels (\emph{e-i}) show how the distribution 
of compartment content, together with their correlations, change over a spectrum 
of different timescale ratios.
We speculate that the approximation (dashed lines) of 
moments $\aesp{M^{1,0}}$, $\aesp{M^{1,1}}$ and $\rho$ (panels \emph{b-d}) 
do not match the Monte Carlo estimation (solid lines) pefectly,
likely because the distributions are more disperse than in the other
case studies we presented in this work 
(see Appendix~\cref{fig:bbdc_histograms,fig:SAIC_histograms}).
Thus, while the approximation is still able to capture the qualitative
dynamics of the system, this case study also reveals the potential limitations
of the approach when dealing with strongly non-linear systems with disperse populations.

\section{Discussion}
Compartmentalization of biochemical reactions into compartments is a hallmark 
of biological systems. 
Studying the dynamic interplay between chemical reactions and compartmental 
dynamics is mathematically challenging, especially in the presence of noise. 
In this work we have developed LPAC, a systematic moment approximation to analyze 
the statistical properties of compartmentalized reaction systems. 
Our approach is based on our earlier work~\cite{Duso2020}, but replaces 
\emph{ad hoc} moment-closure approximations by 
polynomial approximations. 
This renders the approach applicable to a broader class of systems, including 
systems involving non-polynomial rate-laws. 
Our approach exploits the particular structure of the underlying moment equations, 
where individual terms can be written as expectations of products of two functions, 
one depending exclusively on the compartment content and one depending exclusively 
on the number distribution. 
We have further shown that these two functions give rise to two different types 
of closure problems, which is a key difference to bulk chemical systems as well 
as compartmentalized systems in the large $N$ limit. 
The proposed moment-approximation technique addresses both of these closure problems 
and therefore always leads a closed system of moment equations. 
We have included this approach in a new release of our previously developed 
moment generator \emph{Compartor}~\cite{pietzsch2021compartor}, which can be 
used to derive moment equations in a fully automated manner for a broader 
class of systems.

To analyze the performance of our method, we have applied it to several models 
exhibiting nonlinear dynamics. 
We generally found that the approximate moments agreed very well with 
those obtained from exact Monte Carlo simulations. 
This is especially the case for the first two case studies, in which the 
compartment content was narrowly distributed (\Cref{fig:bbdc,fig:SAIC,fig:bbdc_histograms,fig:SAIC_histograms}). 
However, quantitative deviations were observed in situations where the 
distribution of molecular content is strongly disperse (\Cref{fig:MutualRepression,fig:MutualRepressionComparison}). 
This is in line with the fact that a simple truncated Taylor expansion 
was used as a polynomial approximation, 
which is guaranteed to be accurate only locally around the chosen expansion point. 
We remark, however, that our approach could in principle be extended to other 
polynomial approximations.
Polynomial interpolation, for instance, could be used to achieve 
better approximation accuracy across larger domains, 
if a suitable set of interpolation nodes is chosen. 
This in turn could achieve accurate results even in the presence of strongly 
disperse or multimodal distributions. 
Also Bernstein polynomial may provide a promising approximation technique. 
As shown by \emph{Lunz et al.}~\cite{lunz2022revisiting}, they can be 
numerically better behaved than interpolants, although it is currently unclear 
how they perform on the particular class of systems that we considered in this work.
Exploring these techniques more deeply in the context of these systems is an important goal for the future as it may expand the scope of our approach to an even larger class of systems.

\section{Acknowledgements}
We thank Jakob Ruess and Elisa Nerli for their helpful comments on the manuscript.
The authors were supported by core funding of 
the Max Planck Institute of Molecular Cell Biology and Genetics and 
the BMBF under project number 031L0258C with the title 
``In-depth spatial organisation of hepatocellular carcinoma initiation''.

{%
	\footnotesize

\input{0_lpac-paper_arXiv.bbl}
}

\clearpage 
\appendix
\section*{Supplementary Material}
\addcontentsline{toc}{section}{Appendices}

\renewcommand{\thesubsection}{\Alph{subsection}}
\newcommand{\thesubsectionshort}{\Alph{subsection}}
\renewcommand{\theequation}{\thesubsectionshort.\arabic{equation}}
\setcounter{equation}{0}

\subsection{Equations for products of moments} \label{supp:ito}
As mentioned in the main text, it is possible to derive moment equations for generic products of moments.
Let us consider an $m$-ary product of moments 
\begin{equation}
	F(M^{\gamma_1}, \ldots, M^{\gamma_m}) \coloneqq M^{\gamma_1} \ldots M^{\gamma_m}.
\end{equation}
Using It\^o's rule for counting processes~\cite{oksendal2007applied}, it is straightforward to show that the expectation of this moment satisfies the differential equation
\begin{align}
	\begin{split}
		\ddt{} \esp{F(M^{\gamma_1}, &{}\ldots, M^{\gamma_m})} \\
		= \sum_c \besp{
			\sum_{x_c} 
			&{}\bigg[
				F(M^{\gamma_1} + \Delta M_c^{\gamma_1}(x_c), \ldots, M^{\gamma_m} + 
				\Delta M_c^{\gamma_m}(x_c)) \\
				&{} - F(M^{\gamma_1}, \ldots, M^{\gamma_m}) \bigg] \, 
				g_c(x_c) \, w_{c,x_c}(\n)
		}
	\end{split} \\
	\begin{split}
		= \sum_c \besp{
			\sum_{x_c} 
			&{}\bigg[
				\left( M^{\gamma_1} + \Delta M_c^{\gamma_1}(x_c) \right)
				\cdots
				\left( M^{\gamma_m} + \Delta M_c^{\gamma_m}(x_c) \right) \\
			&{} - M^{\gamma_1} \, \ldots\, M^{\gamma_m} \bigg] \, 
				g_c(x_c) \, w_{c,x_c}(\n)
		} \;.
	\end{split} \label{eq:itoProdExpanded}
\end{align}

We can expand the products of all binomials in~\cref{eq:itoProdExpanded} and 
get a sum of products of moments and $\Delta M_c^{\gamma_j}$-terms. Each of these products
contains, at most, $m-1$ moments.
\begin{remark}
	Each term in the fully expanded version of~\cref{eq:itoProdExpanded} has 
	the form
	\begin{equation}
		\prod_{k\in\K} M^{\gamma_k} 
		\, \underbrace{
			\prod_{h\in\H} \Delta M_c^{\gamma_h}(x_c)
			\, g_c(x_c)
			}_{
			f_c^\H(x_c)
			}
		\, w_{c,x_c}(\n) \;,
		\label{eq:itoTerm}
	\end{equation}
	where if we define $\Gamma \coloneqq \{\gamma_1,\ldots,\gamma_m\}$,
	then $\K \subset \Gamma$ s.t. $|\K| \leq m-1$
	and $\H = \Gamma \setminus \K$.
\end{remark}

By substituting the definition of a population moment into~\cref{eq:itoTerm}
we get that each term in~\cref{eq:itoProdExpanded} can be written as follows:
\begin{align}
	\bigg[ {}& \prod_{k\in\K}
		\sum_{i=1}^{|\K|} (x_\K^{(i)})^{\gamma_k} \, \n_{x_\K^{(i)}} \bigg]
		\, f_c^\H(x_c)
		\, w_{c,x_c}(\n)
		\notag \\
	={}& \sum_{x_\K} g^\K(x_\K) \, w_\K(\n; x_\K)
		\, f_c^\H(x_c) \, w_{c,x_c}(\n)
		\notag \\
	={}& \sum_{x_\K} f_c^{\H,\K}(x_c, x_\K) \, w_{c,\K}(\n; x_c, x_\K) \;.
\end{align}
Here $x_\K$ is a tuple of compartment content symbols involved in the cross
product identified by the set $\K$, 
in the same way as $x_c$ represents the tuple of reactant compartments of a 
given transition $c$.
\enlargethispage{\baselineskip} 

Now we can rewrite the whole moment equation for a generic product of moments as:
\begin{align}
	\ddt{} {}& \aesp{M^{\gamma_1} \, \ldots\, M^{\gamma_m}} \notag \\
	={}& \sum_c \besp{
		\sum_{x_c} \sum_\K \sum_{x_\K} 
			f_c^{\H,\K}(x_c, x_\K) \, w_{c,\K}(\n; x_c, x_\K)
		}
		\notag \\
	={}& \sum_c \sum_\K \besp{
		\sum_{(x_c, x_\K)} f_c^{\H,\K}(x_c, x_\K) \, w_{c,\K}(\n; x_c, x_\K)
		} \;.
\end{align}
This leads us directly to the following:
\begin{lemma}
	Computing the contribution of a $r_c$-compartmental transition to the equation
	of a product of $m$ moments is equivalent to a ``virtual'' transition involving
	$r_c + m - 1$ compartments, with a propensity that is still separable into
	a content-dependent function $f$ and a state-dependent function $w$.
\end{lemma}
\begin{remark}
	This means that all the results on approximating moment equations
	of the form
	\begin{equation*}
		\ddt{\aesp{M^\gamma}}
	\end{equation*}
	can also be applied to any arbitrary product of 
	moments
	\begin{equation*}
		\ddt{}\aesp{M^{\gamma_1} \, \ldots\, M^{\gamma_m}}
	\end{equation*}
	and therefore to derive the equation for the expectation of any arbitrary
	moment polynomial.
\end{remark}

\subsection{Approximated moment equations} \label{supp:approx}
Let us now consider the separated form of a generic moment 
equation~\eqref{eq:separatedDiffeq} and assume that all content and state 
functions have been approximated by suitable polynomials.
Expanding the products leads to monomials of the form
\begin{equation}
	C \, 
	x_{c,1,1}^{\alpha_{c,1,1}} \cdots x_{c,r_c,d}^{\alpha_{c,r_c,d}} \, 
	\n_{x_{c,1}}^{\beta_{c,1}} \cdots \n_{x_{c,r_c}}^{\beta_{c,r_c}} \;,
	\label{eq:expansionMonomial}
\end{equation}
where $C$ contains all the constant terms of the expression.

Since the expectation and summation $\aesp{ \sum_{x_c} \cdots }$ are linear, we
can bring both of them in and around each monomial. With a rearrangement of the 
operands and then splitting the summation we get:
\begin{align}
	\begin{split} \notag
	\besp{
		\sum_{x_c}
			C \, 
			&{} x_{c,1,1}^{\alpha_{c,1,1}} \cdots x_{c,1,d}^{\alpha_{c,1,d}} 
				\, \n_{x_{c,1}}^{\beta_{c,1}} \\
			&{} \, \ldots \,
			x_{c,r_c,1}^{\alpha_{c,r_c,1}} \cdots x_{c,r_c,d}^{\alpha_{c,r_c,d}} 
				\, \n_{x_{c,r_c}}^{\beta_{c,r_c}}
	} 
	\end{split} \\
	\begin{split}
	= C \, \besp{
			&{} \sum_{x_{c,1}} x_{c,1,1}^{\alpha_{c,1,1}} \cdots x_{c,1,d}^{\alpha_{c,1,d}} 
				\, \n_{x_{c,1}}^{\beta_{c,1}} \\
			&{} \, \ldots \,
			\sum_{x_{c,r_c}} x_{c,r_c,1}^{\alpha_{c,r_c,1}} \cdots x_{c,r_c,d}^{\alpha_{c,r_c,d}} 
				\, \n_{x_{c,r_c}}^{\beta_{c,r_c}}
	} \label{eq:expansionMonomialMomentExpanded} 
	\end{split} \\
	\begin{split}
	= C \, \besp{
			&{} M^{\alpha_{c,1}, \beta_{c,1}} 
			\, \ldots \,
			M^{\alpha_{c,r_c}, \beta_{c,r_c}}
	} \;. \label{eq:expansionMonomialMoment}
	\end{split}
\end{align}
\begin{notation}
	Here we have used an extended notation for generalized population moments:
	\begin{equation}
		M^{\gamma,\sigma} \coloneqq \sum_x x^\gamma \n_x^\sigma \;.
	\end{equation}
	With the term \emph{order} we still refer to $|\gamma|$, while we refer to
	$|\sigma|$ as \emph{\norder{}}.
\end{notation}
\begin{remark}\label{remark:supp:wMultilinear}
	If the compartment events follow mass-action-like kinetics, then their state
	function $w$ is multilinear, i.e. it is linear with respect to each single reactant compartment symbol.
	This means that it can only generate moments that have a maximum \norder{} 
	of 1 each.
\end{remark}

\begin{lemma} \label{lemma:orderIsBound}
	If the polynomial approximating the content function $f$ has degree $\delta$,
	then for any moment $M^{\gamma,\sigma}$ that appears in the approximated 
	moment equation it holds $|\gamma| \leq \delta$.
\end{lemma}
\begin{proof}
	Since the degree of the polynomial is $\delta$, then by definition the
	sum of the $\alpha$ exponents of all monomials in~\cref{eq:expansionMonomial} 
	is bounded by $\delta$.
	This means that the order of all moments generated by these monomials is 
	also necessarily bounded by $\delta$.
\end{proof}
\begin{remark}
	From the proof we also get a stronger result: the total order of the moments
	in any moment product is also bounded by the approximation degree $\delta$.
\end{remark}

\begin{lemma} \label{lemma:norderIsBound}
	If the polynomial approximating the content function $w$ has degree $\varepsilon$, 
	then the total \norder{} of any moment product
	\begin{equation}
		\aesp{
			M^{\gamma_1,\sigma_1} \, \ldots \, M^{\gamma_m,\sigma_m}
		}
	\end{equation}
	is bounded by $\varepsilon$, that is $\sum_{i=1}^m \sigma_i \leq \varepsilon$.
\end{lemma}
\begin{proof}
	Since the degree of the polynomial approximating $w$ is $\varepsilon$, 
	then all monomials in~\cref{eq:expansionMonomial} have a total order
	of the $\beta$ exponents that is bounded by $\varepsilon$.
	Following the derivation of~\cref{eq:expansionMonomialMomentExpanded} we 
	then obtain that any product of moments obtained by the approximation and 
	truncation procedure has a total \norder{} bounded by $\varepsilon$.
\end{proof}
\begin{remark}
	This also means that the number of moments involved in any product appearing
	on the right-hand side is also bounded. In case the compartment events follow
	a mass-action-like kinetics, then the number of moments must be $\leq \varepsilon$.
\end{remark}

The above results lead to the following:
\begin{theorem}
	Given an equation for any arbitrary moment or product of moments and two
	approximation degrees $\delta, \varepsilon \in \N$, the approximation
	of order $(\delta,\varepsilon)$ of such equation can be solved through a
	system of moment equations with the same approximation order and the system 
	is closed.
\end{theorem}
\begin{proof}
	Using~\cref{lemma:orderIsBound,lemma:norderIsBound} we obtain that the equation
	for any desired moment, approximated at order $(\delta,\varepsilon)$ only
	contains products of moments of total order $\delta$ and total \norder{}
	$\varepsilon$. Furthermore any product of moments additionally required in
	the system of equations can also be described with an approximated equation
	only containing products of moments of total order $\delta$ and total \norder{}
	$\varepsilon$.
	Since order and \norder{} are positive integers and bounded by $\delta$ and 
	$\varepsilon$ respectively, then the combinations of moments that can
	appear in a product of moments are finite and this proves that the system of
	equations is closed.
\end{proof}

We therefore have a simple algorithm allowing to always derive a closed set of
moment equations: first derive the equations for all the moments of interest,
substituting all content- and state functions with suitable polynomials of limited
degree; then collect all moments and moment products these equations depend
on and iteratively apply the same procedure for all those we still have no 
equation for. This procedure will terminate and produce a closed set of moment equations.

\subsection{Multi-index notation and operators} \label{supp:multi-index}
Throughout this work, we rely on standard multi-index notation to make 
the equations more compact and easier to read.
Multi-index notation allows common operators that accept integer indices as arguments
to be extended to ordered tuples of indices.
Here we provide the definitions of multi-index operators used in our work.

A $d$-dimensional multi-index is defined as a $d$-tuple:
\begin{equation*}
	\alpha = (\alpha_1, \alpha_2, \ldots, \alpha_d) \in \N_0^d \;.
\end{equation*}
If $\alpha, \beta \in \N_0^d$ and $x \in \R^d$ we can define:
\paragraph{Sum}
	\begin{equation*}
		\alpha + \beta = (\alpha_1 + \beta_1, \ldots, \alpha_d + \beta_d)
	\end{equation*}
\paragraph{Partial ordering}
	\begin{equation*}
		\alpha \leq \beta = (\alpha_1 \leq \beta_1, \ldots, \alpha_d \leq \beta_d)
	\end{equation*}
\paragraph{Power}
	\begin{equation*}
		x^\alpha = x_1^{\alpha_1} \cdots x_d^{\alpha_d}
	\end{equation*}
\paragraph{Factorial}
	\begin{equation*}
		\alpha! = \alpha_1! \cdots \alpha_d!
	\end{equation*}
\paragraph{Binomial coefficient}
	\begin{equation*}
		\binom{\alpha}{\beta} = \binom{\alpha_1}{\beta_1} \cdots \binom{\alpha_d}{\beta_d}
	\end{equation*}
\paragraph{Partial derivatives}
	\begin{equation*}
		\D^\alpha = \frac{\partial^{\alpha_1}}{\partial x_1^{\alpha_1}}
			\cdots  \frac{\partial^{\alpha_d}}{\partial x_d^{\alpha_d}}
	\end{equation*}

\subsection{Including bulk chemical species and reactions} \label{supp:bulk}
It is possible to have systems where a compartment population, with its internal
chemistry, interacts with chemistry occuring in the bulk of the outer medium.
We gave an example coming from bio-engineering in~\cref{sec:saic}, but many other biological systems have this characteristic too.

Here we want to show how the same mathematical framework that we used can
be extended in order to account for bulk chemical species and how their
chemistry can be coupled to the compartmentalised system and its transitions.

Let us recall the generic 3-term form of a moment equation, 
from~\cref{eq:applicableDiffeq}:
\begin{equation*}
	\ddt{\esp{M^\gamma}} = \sum_c
	\besp{ 
		\sum_{x_c}
		\Delta M_c^\gamma(x_c) \, g_c(x_c)
		\, w_{c,x_c}(\n)
	} \;.
\end{equation*}
We introduce a new stochastic state vector $\B$, which collects the copy numbers of all
bulk chemical species. 
We can now account for the bulk chemical state in the moment equations by 
assuming the transition propensities also have a dependence on the bulk state:
\begin{equation}
\begin{split}
	\ddt{\esp{M^\gamma}} = \sum_c
	\besp{ 
		\sum_{x_c}
		\Delta M_c^\gamma(x_c) \, g_c(x_c) 
		\, w_{c,x_c}(\n) \, b_c(\B)
	} \;.
\end{split}
\end{equation}
Where the \emph{bulk-dependent} rate function $b_c$ would have a constant value 
of $1$ in each transition class that does not depend on the bulk chemical state.

Analogously to the presented moment equations, we can also derive differential equations for 
the evolution of the expected bulk state:
\begin{equation}
\begin{split}
	\ddt{\esp{\B}} = \sum_c
	\besp{ 
		\sum_{x_c}
		\Delta B_c(x_c) \, g_c(x_c) 
		\, w_{c,x_c}(\n) \, b_c(\B)
	} \;.
\end{split} \label{suppeq:bulkMomentEquation}
\end{equation}
Chemical reactions that occur in the bulk, without involving compartments and
their content, are represented by transition classes that do not
involve any reactant compartment, i.e. with $x_c = \{\}$.
In this case the corresponding terms in~\cref{suppeq:bulkMomentEquation}
contain a $\Delta B_c$ that is independent of $x_c$ 
such that the inner sum simplifies to
$\Delta B_c
	\aesp{ 
		b_c(\B)
	}
$.
In the special case where there are no chemical interactions between any bulk and 
compartmentalised chemical species, \cref{suppeq:bulkMomentEquation} 
would further simplify to a conventional moment equation for bulk chemical systems
\begin{equation}
	\ddt{\esp{\B}} = \sum_c
	\Delta B_c \, 
	\aesp{ 
		b_c(\B)
	} \;.
\end{equation}

The function of the bulk state $b_c(\B)$ can then be polynomially expanded
around the mean state $\esp{\B}$ in the same way as the state function 
$w_{c,x_c}(\n)$, obtaining a closed set of moment equations that account for bulk
quantities and moment-bulk cross products.

\subsection{Pearson correlation coefficient in compartment populations} \label{supp:correlation}
We use a mean-field approximation of the Pearson correlation coefficient in order
to be able to compute it from the population moments that are already available
in the \emph{Mutual Repression} case study from~\cref{sec:MutualRepression}.

We use the following base approximations:
\begin{align}
	\cov(x_1,x_2) &{}= \aesp{\frac{M^{1,1}}{N}} - \aesp{\frac{M^{1,0}}{N}} \, \aesp{\frac{M^{0,1}}{N}} \notag \\
	&{}\approx \frac{\aesp{M^{1,1}}}{\aesp{N}} - \frac{\aesp{M^{1,0}} \, \aesp{M^{0,1}}}{\aesp{N}^2} \\
	\var(x_1) &{}= \aesp{\frac{M^{2,0}}{N}} - \aesp{\frac{M^{1,0}}{N}}^2 \notag \\
	&{}\approx \frac{\aesp{M^{2,0}}}{\aesp{N}} - \frac{\aesp{M^{1,0}}^2}{\aesp{N}^2} \;.
\end{align}
Plugging them into the Pearson correlation's definition we get the formula that 
we use in the analysis of the case study:
\begin{align}
	\rho(x_1,x_2) &{}= \frac{ \cov(x_1,x_2) }{ \sqrt{\var(x_1)} \, \sqrt{\var(x_2)} } \notag \\
	&{}\approx \frac{ \aesp{M^{1,1}} - \frac{ \aesp{M^{1,0}} \, \aesp{M^{0,1}} }{ \aesp{N} } }%
	{ \sqrt{ \aesp{M^{2,0}} - \frac{\aesp{M^{1,0}}^2}{\aesp{N}} } \, \sqrt{ \aesp{M^{0,2}} - \frac{\aesp{M^{0,1}}^2}{\aesp{N}} } } \;.
\end{align}

\subsection{Case study: Binary birth-death-fusion process} \label{supp:bbdc}
\begin{figure}[t]
	\centering
	\includegraphics[scale=\figscale]{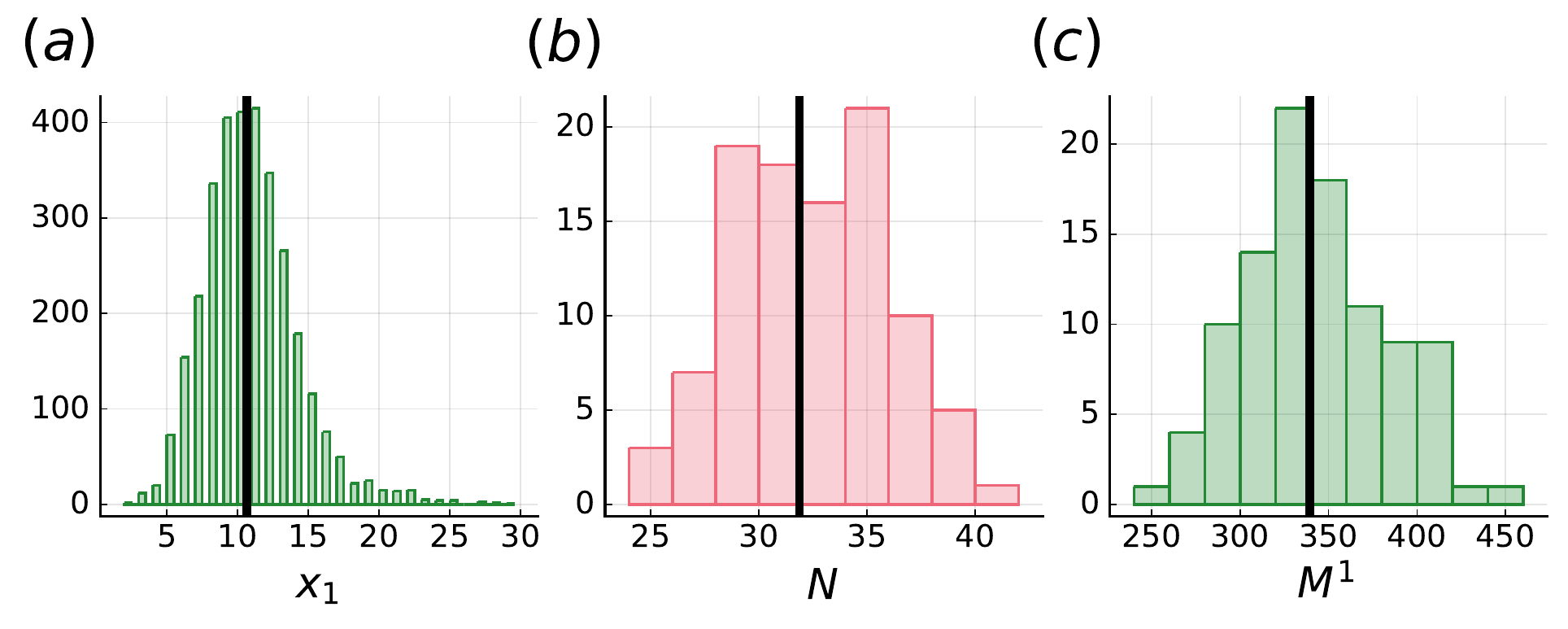}%
	\caption{
	Histograms for the distributions of 
	(\emph{a})~the cellular content variable $x_1$, 
	(\emph{b})~the number of cells in the system $N$ and 
	(\emph{c})~the total mass of $x_1$ in the system $M^1$
	in the \emph{Binary birth-death-fusion} case study.
	The solid black lines in each panel shows the estimated values of
	$\aesp{M^1}/\aesp{N}$, $\aesp{N}$ and $\aesp{M^1}$, respectively, as computed
	by the approximation.
	}%
	\label{fig:bbdc_histograms}
\end{figure}

As detailed in~\cref{sec:bbdc} of the main text, 
the system is defined by the transition classes
\begin{align}
	\emptyset &\react{h_I(\n; y)} [y] && \text{(Intake)} \notag \\
	[x] + [x'] &\react{h_F(\n; x,x')} [x + x'] && \text{(Fusion)} \notag \\
	[x] &\react{h_b(\n; x)} [x+1] && \text{(Birth)} \notag \\
	[x] &\react{h_d(\n; x)} [x-2] \;. && \text{(Death)} \notag
\end{align}
with propensity functions
\begin{align}
	h_I(\n; y) &= k_I \, \pi_I(y) \label{suppeq:bbdc:hI} \\
	h_F(\n; x,x') &= k_F \, \frac{\n_x \, (\n_{x'} - \delta_{x,x'})}{1 + \delta_{x,x'}} \label{suppeq:bbdc:hE} \\
	h_b(\n; x) &= k_b \, \n_x \label{suppeq:bbdc:hb} \\
	h_d(\n; x) &= k_d \, \frac{x \, (x-1)}{2} \, \n_x \;. \label{suppeq:bbdc:hd} 
\end{align}

The complete set of moment equations for this case study,
together with the code to generate them automatically with Compartor, 
can be found in the corresponding Jupyter notebook in the
~\href{https://github.com/zechnerlab/Compartor/blob/master/(LPAC)%201%20Binary%20Birth-Death-Fusion.ipynb}{Compartor repository} on GitHub.

\subsection{Case study: Shared Antithetic Integral Controller} \label{supp:SAIC}
\begin{figure*}[t]
	\centering
	\centerline{
		\includegraphics[scale=\figscale]{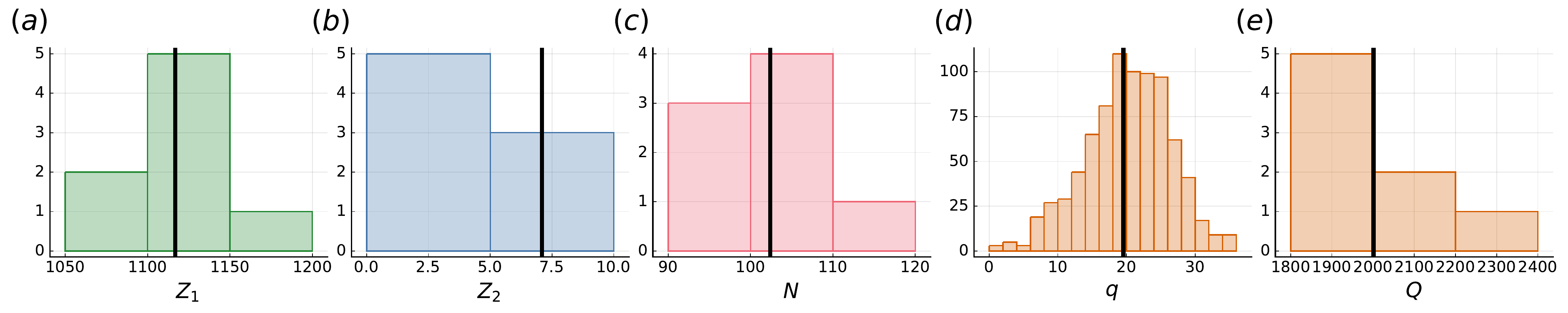}}%
	\caption{
	Histograms for the distributions of (\emph{a},\emph{b})~the bulk controller
	species $Z_1$ and $Z_2$ respectively, 
	(\emph{c})~the number of cells in the system $N$,
	(\emph{d})~the cellular content variable $q$ and 
	(\emph{e})~the total mass of $q$ in the system $Q = M^1$
	in the \emph{Shared Antithetic Integral Controller} case study.
	The solid black lines in each panel shows the estimated values of
	$\aesp{Z_1}$, $\aesp{Z_2}$, $\aesp{N}$, 
	$\aesp{Q}/\aesp{N}$ and $\aesp{Q}$, respectively, as computed
	by the approximation.
	}%
	\label{fig:SAIC_histograms}
\end{figure*}

The \emph{sAIC} case study features a bulk chemistry that is coupled to the
compartmentalised system. 
As detailed in Supplementary Material~\cref{supp:bulk}, bulk chemistry can be integrated into the 
mathematical framework with little symbolic overhead.
\emph{Compartor}, our tool for generating the moment equations, does however not
yet support this extension.
We have therefore resorted to ``virtualising'' the sAIC bulk chemistry into a set
of compartmentalised transitions that have identical dynamics.

The internal compartment state is given by the 3-species tuple $(Z_1, Z_2, Q)$,
where $Z_1$ and $Z_2$ are the two bulk controller chemical species and $Q$ is the
actual internal measured species.
Species $Z_1$ and $Z_2$ are therefore virtually spread among the compartments in
the system and their total amounts are given by the moments $M^{1,0,0}$ and 
$M^{0,1,0}$ respectively.
The transition classes are modified in order to conserve $Z_1$ and $Z_2$ in case
of compartmental events.
Furthermore, the \emph{Comparison} reaction is split into two transitions, in 
order to account for the reactant molecules to virtually belong either to the
same compartment or to two different ones.

\enlargethispage{\baselineskip} 
The full augmented transition network is represented by 
the following diagram:
\begin{align}
	[x] &\react{h_D(\n; x,y)} [y] + [x-y] & \text{(Division)} \notag \\
	[x] + [x'] &\react{h_A(\n; x)} [x + (x'_1,x'_2,0)] & \text{(Apoptosis)} \notag \\
	[x] &\react{h_p(\n; x)} [x + (0,0,1)] & \text{(Production)} \notag \\
	[x] &\react{h_d(\n; x)} [x - (0,0,1)] & \text{(Degradation)} \notag \\
	[x] &\react{h_{ref}(\n; x)} [x + (1,0,0)] & \text{(Reference)} \notag \\
	[x] &\react{h_{act}(\n; x)} [x + (0,0,1)] & \text{(Actuation)} \notag \\
	[x] &\react{h_{meas}(\n; x)} [x + (0,1,0)] & \text{(Measurement)} \notag \\
	[x] &\react{h_{comp_1}(\n; x)} [x + (-1,-1,0)] & \text{(Comparison 1)} \notag \\
	[x] + [x'] &\react{h_{comp_2}(\n; x,x')} [x + (-1,0,0)] \notag \\
	& \phantom{\react{h_{comp_2}(\n; x,x')}} + [x' + (0,-1,0)] \;, & \text{(Comparison 2)} \notag
\end{align}
while the propensity functions are as follows:
\begin{align}
	h_D(\n; x,y) &= k_D \, \pi_D(y \mid x) \, x_3 \, \n_x \label{suppeq:saic:hD} \\
	h_A(\n; x,x') &= k_A \, \frac{\n_x (\n_{x'} - \delta_{x,x'})}{1 + \delta_{x,x'}} \label{suppeq:saic:hA} \\
	h_p(\n; x) &= k_p \, \n_x \label{suppeq:saic:hp} \\
	h_d(\n; x) &= k_d \, x_3 \, \n_x \label{suppeq:saic:hd} \\
	h_{ref}(\n; x) &= \frac{k_{ref}}{N} \, \n_x \label{suppeq:saic:hRef} \\
	h_{act}(\n; x) &= k_{act} \, M^{1,0,0} \, \n_x \label{suppeq:saic:hAct} \\
	h_{meas}(\n; x) &= k_{meas} \, x_3 \, \n_x \label{suppeq:saic:hMeas} \\
	h_{comp_1}(\n; x) &= k_{comp_1} \, x_1 x_2 \, \n_x \label{suppeq:saic:hComp1} \\
	h_{comp_2}(\n; x,x') &= k_{comp_2} \, x_1 x'_2 \, \frac{\n_x (\n_{x'} - \delta_{x,x'})}{1 + \delta_{x,x'}} \;. \label{suppeq:saic:hComp2}
\end{align}

The complete set of moment equations for this case study,
together with the code to generate them automatically with Compartor, 
can be found in the corresponding Jupyter notebook in the
~\href{https://github.com/zechnerlab/Compartor/blob/master/(LPAC)%202%20Shared%20Antithetic%20Integral%20Controller.ipynb}{Compartor repository} on GitHub.

\subsection{Case study: Mutually repressing gene circuit in a cell population} \label{supp:MutualRepression}
\begin{figure*}[p] 
	\centering
	\vspace*{-2cm}
	\centerline{
		\includegraphics[scale=\widefigscale]{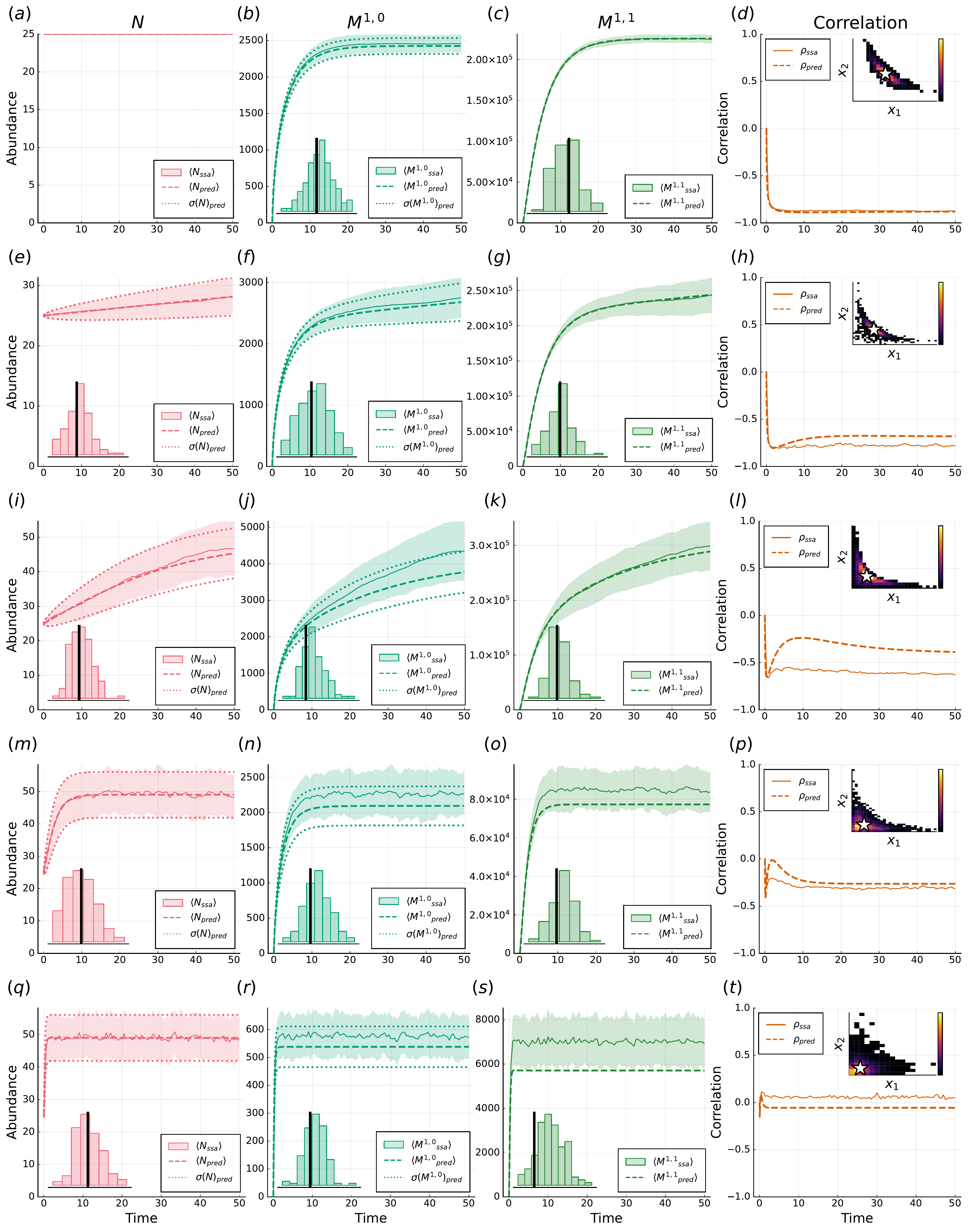}}%
	\caption{
	Comparison of the results obtained by averaging 128 SSA trajectories, 
	shown by the solid line and shaded area,
	with the predictions of our moment-expansion method, shown by the dashed line
	and dotted boundaries. The solid and dashed lines denote average values, while
	the shaded and dot-bordered areas are the regions within one standard deviation.
	The layout of the panels for each row follows that of (\cref{fig:MutualRepression}, \emph{a-d}),
	while the rows correspond to the different $k_b/k_D$ regimes as in  
	(\cref{fig:MutualRepression}, \emph{e-i}):
	(\emph{a-d}) $k_b/k_D = \infty$,
	(\emph{e-h}) $k_b/k_D = 20000$,
	(\emph{i-l}) $k_b/k_D = 2000$,
	(\emph{m-p}) $k_b/k_D = 200$,
	(\emph{q-t}) $k_b/k_D = 20$.
	}
	\label{fig:MutualRepressionComparison}
\end{figure*}

This case study features two internal chemical species, representing levels of
gene expression, each inhibiting the production of the other through a repressor
Michaelis-Menten type of kinetics.
This internal dynamics is then imposed on a population of dividing cells, which 
size is kept in check by a binary cell death, i.e. a bi-compartmental transition
in which one of the reactant compartments leaves the population.

The full transition network is represented by 
the following diagram:
\begin{align}
	[x] &\react{h_D(\n; x,y)} [y] + [x-y] & \text{(Division)} \notag \\
	[x] + [x'] &\react{h_A(\n; x)} [x] & \text{(Apoptosis)} \notag \\
	[x] &\react{h_{p_1}(\n; x)} [x + (1,0)] & \text{(Production 1)} \notag \\
	[x] &\react{h_{d_1}(\n; x)} [x - (1,0)] & \text{(Degradation 1)} \notag \\
	[x] &\react{h_{p_2}(\n; x)} [x + (0,1)] & \text{(Production 2)} \notag \\
	[x] &\react{h_{d_2}(\n; x)} [x - (0,1)] \;. & \text{(Degradation 2)} \notag
\end{align}
While the propensity functions are as follows:
\begin{align}
	h_D(\n; x,y) &= k_D \, \pi_D(y \mid x) \, \n_x \label{suppeq:MR:hD} \\
	h_A(\n; x,x') &= k_A \, \frac{\n_x (\n_{x'} - \delta_{x,x'})}{1 + \delta_{x,x'}} \label{suppeq:MR:hA} \\
	h_{p_1}(\n; x) &= k_p \, \frac{k_{R_1}}{k_{R_1} + x_2} \, \n_x \label{suppeq:MR:hp1} \\
	h_{d_1}(\n; x) &= k_d \, x_1 \, \n_x \label{suppeq:MR:hd1} \\
 	h_{p_2}(\n; x) &= k_p \, \frac{k_{R_2}}{k_{R_2} + x_1} \, \n_x \label{suppeq:MR:hp2} \\
	h_{d_2}(\n; x) &= k_d \, x_2 \, \n_x \;. \label{suppeq:MR:hd2} 
\end{align}

The complete set of moment equations for this case study,
together with the code to generate them automatically with Compartor, 
can be found in the corresponding Jupyter notebook in the
~\href{https://github.com/zechnerlab/Compartor/blob/master/(LPAC)%203%20Mutual%20Repression.ipynb}{Compartor repository} on GitHub.


\end{document}

%% file: 1_definitions.tex
\newcommand{\figscale}{0.42}
\newcommand{\widefigscale}{0.42}

\newtheorem{lemma}{Lemma}
\newtheorem{theorem}{Theorem}

\newtheorem{remark}{Remark}
\newtheorem*{notation}{Notation}
\newcommand*{\react}[2][]{%
	\ifthenelse{\equal{#1}{}}%
	{\xrightarrow{#2}}%
	{\xrightarrow{\mathmakebox[#1]{#2}}}%
}

\newcommand*{\bireact}[3][]{%
	\ifthenelse{\equal{#1}{}}%
	{\xrightleftharpoons[#3]{#2}}%
	{\xrightleftharpoons[#3]{\mathmakebox[#1]{#2}}}%
}

\newcommand{\R}{\mathbb{R}}
\newcommand{\X}{\mathbb{X}}
\newcommand{\N}{\mathbb{N}}
\newcommand{\E}{\mathbb{E}}

\newcommand{\n}{\bm{n}}
\newcommand{\B}{\bm{B}}
\newcommand{\e}{\bm{e}}

\newcommand{\D}{\mathcal{D}}
\renewcommand{\H}{\mathcal{H}}
\newcommand{\K}{\mathcal{K}}
\newcommand{\C}{\mathcal{C}}

\newcommand{\bmD}{\bm{\D}}
\newcommand{\bmeta}{\bm{\eta}}
\newcommand{\de}{\mathrm{d}}
\newcommand*{\ddt}[1]{\frac{\de #1}{\de t}}
\newcommand*{\esp}[2][]{%
	\ifthenelse{\equal{#1}{}}%
	{\langle #2 \rangle}%
	{\langle #2 \rangle_{#1}}%
}
\newcommand*{\aesp}[2][]{%
	\ifthenelse{\equal{#1}{}}%
	{\left\langle #2 \right\rangle}%
	{\left\langle #2 \right\rangle_{#1}}%
}
\newcommand*{\besp}[2][]{%
	\ifthenelse{\equal{#1}{}}%
	{\Big\langle #2 \Big\rangle}%
	{\Big\langle #2 \Big\rangle_{#1}}%
}
\newcommand*{\Esp}[2][]{%
	\ifthenelse{\equal{#1}{}}%
	{\E\!\left[ #2 \right]}%
	{\E_{#1}\!\left[ #2 \right]}%
}

\newcommand*{\norder}{$\n$-order}

\newcommand{\var}{\mathrm{var}}
\newcommand{\cov}{\mathrm{cov}}